\documentstyle[12pt]{article}

\newcommand{\T}{T_{\mbox{\scriptsize vac}}^{\mu\nu}}
\renewcommand{\t}{T_{\mbox{\scriptsize source}}^{\mu\nu}}
\renewcommand{\O}{{\cal O}}
\renewcommand{\u}{u=\mbox{const.}}
\renewcommand{\v}{v=\mbox{const.}}
\newcommand{\rr}{(\nabla r)^2}
\newcommand{\rv}{(\nabla r,\nabla v)}
\newcommand{\ru}{(\nabla r,\nabla u)}
\newcommand{\uv}{(\nabla u,\nabla v)}
\newcommand{\dr}{\frac{\partial r}{\partial u}}
\newcommand{\drplus}{\frac{\partial r^{\hphantom{+}}}{\partial u^+}}
\newcommand{\strong}{-\frac{\partial r^{\hphantom{+}}}{\partial u^+}
={\cal O}}
\newcommand{\weak}{-\frac{\partial r^{\hphantom{+}}}{\partial u^+}
>O(\sqrt{{\bf\cal A}})}
\newcommand{\rH}{r_{\mbox{\tiny AH}}}
\newcommand{\EH}{E_{\mbox{\tiny AH}}}
\newcommand{\BH}{B_{\mbox{\tiny AH}}}
\newcommand{\uH}{u_{\mbox{\tiny AH}}}
\newcommand{\vH}{v_{\mbox{\tiny AH}}}
\newcommand{\I}{{\cal I}^+}
\newcommand{\mI}{{\cal I}^-}
\renewcommand{\|}{\biggl |_{{\cal I}^+}}
\newcommand{\scm}{\biggl |_{{\cal I}^-}}
\makeatletter
\@addtoreset{equation}{section}
\makeatother
\begin{document}

{\renewcommand{\theequation}{1.\arabic{equation}}

\begin{center}
{\LARGE\bf
Kinematics}\\{\LARGE\bf
of evaporating black holes }\\
\end{center}
\begin{center}
{\bf G.A. Vilkovisky}
\end{center}
\begin{center}
Lebedev Physical Institute, and Lebedev
Research Center in Physics,\\ Leninsky Prospect 53, 119991 Moscow,
Russia.\\ E-mail: vilkov@sci.lebedev.ru \\
\end{center}
\vspace{3cm}
\begin{abstract}

The correspondence principle and causality divide the spacetime
of a macroscopic collapsing mass into three regions: classical,
semiclassical, and ultraviolet. The semiclassical region covers
the entire evolution of the black hole from the macroscopic to
the microscopic scale if the latter is reached. It is shown that
the metric in the semiclassical region is expressed purely
kinematically through the Bondi charges. The only quantum
calculation needed is the one of radiation at infinity. The
ultraviolet ignorance of semiclassical theory is irrelevant.
The metric with arbitrary Bondi charges is obtained and studied.
\end{abstract}
\newpage

\begin{center}
\section{\bf    Introduction}
\end{center}

$$ $$

The problem of backreaction of the Hawking radiation [1] remains
unsolved. In the general setting, it consists in obtaining the
solution of the expectation-value equations
\begin{equation}
R^{\mu\nu}-\frac{1}{2}g^{\mu\nu}R=8\pi T^{\mu\nu}\;,\qquad
T^{\mu\nu}=\t +\T
\end{equation}
for the gravitational field with the collapse initial conditions.
Here $\t$ is the energy-momentum tensor of a given matter source,
and $\T$ is a certain retarded functional of the curvature: the
energy-momentum tensor of the in-vacuum [2]. Although there has
been a number of studies (for their discussion and the list of
references see the book [3]), no consistent approach to the problem
was proposed. Meanwhile, the Hawking effect is within well-established
physics and is unambigously described by semiclassical field theory.
Semiclassical theory is incapable of giving the full solution of
eq. (1.1) because it fails at small scale but, if the matter source has
macroscopic parameters, there is a region of the expectation-value
spacetime in which semiclassical theory is valid. Moreover, this
region is causally complete (see below) and covers the entire
evolution of the black hole from the macroscopic to the microscopic
scale if the latter is reached. This is the first thing shown in
the present paper. The solution of the backreaction problem
should, therefore, be derivable from semiclassical field theory.

Immediately, two questions emerge. First, the semiclassical $\T$
is a sum of the vacuum loops and is not calculable exactly.
Then what approximations are to be used? Second, in semiclassical
theory, $\T$ is even theoretically calculable only up to terms
local in the curvature and proportional to the quantum constant.
This arbitrariness is the expression of the ultraviolet ignorance
of semiclassical theory. Owing to their locality, the indefinite
terms of $\T$ do not affect the radiation at infinity [2] but how
can it be that they do not affect the metric in the compact domain?
The present paper gives an economical answer to both questions: the
semiclassical $\T$ need not be calculated in the compact domain.
Up to negligible corrections, the metric in the semiclassical region
is expressed through the Bondi charges by a set of kinematic equations.
These equations will close as soon as the Bondi charges will be
expressed through the metric in the semiclassical region. For that,
$\T$ needs to be calculated at infinity and nowhere else. Thereby,
the "backreaction of radiation" acquires a literal meaning. The
indefinite terms of $\T$ do not affect the metric in the semiclassical
region because, in this region, they are negligible altogether.
The latter conclusion conforms to the significance of these terms
and is conditioned by their two distinctive properties: locality,
and proportionality to the quantum constant.

In brief outline, the contents of the paper is seen from the titles
of its sections: (2)~Correspondence principle; (3)~Correspondence
principle (continued); (4)~Solution in the region of strong field;
(5)~Field equations; (6)~Solution in the region of weak field;
(7)~Global solution; (8)~Global solution (continued); (9)~Summary.
The correspondence principle is considered carefully in sections
2 and 3 because it provides the initial conditions for the
expectation-value equations and gives a key to establishing the
bounds of the semiclassical region. The concept of strong field
is introduced in section 4, and it is shown that, in the region of
strong field, the gravity equations close purely kinematically
leaving the arbitrariness only in the data functions. For the
equations to close in the region of weak field, it suffices that
at least one component of $\T$ in a certain basis remain microscopic.
This is discussed in detail in section 5. The solution in the region
of weak field is obtained in section 6 in terms of arbitrary data
functions. The solutions in the regions of strong and weak field
are next sewn together, and the data functions for these regions
are related. In this way in sections 7 and 8 the global solution is
obtained containing two arbitrary data functions. These functions are
the Bondi charges at the future null infinity.

It is assumed below that the matter source in eq. (1.1) has a compact
spatial support and is spherically symmetric. The correspondence
principle will make it possible to limit the consideration to the
vacuum region. Then the only relevant parameter of the source is its
mass $M$ which is also the ADM mass of the expectation-value
spacetime [2]. The principal condition assumed in the present study is
\begin{equation}
\lambda\ll 1
\end{equation}
where
\begin{equation}
\lambda=\frac{\mu}{M}\; ,
\end{equation}
and $\mu$ is the Planckian mass. An observable that in the units
of $M$ (to the appropriate power) vanishes as $\lambda\to 0$ will
be denoted as $\O$. The dimension of $\O$ in terms of $M$ may or
may not be pointed out explicitly. With this notation, inequalities
of the form $X>|\O|$ assume {\it any} $\O$, and equalities of the
form $Y=\O$ assume {\it some} $\O$. Since $\lambda$ is the only
quantum parameter in the problem, these specifications signify
that $X$ is a macroscopic quantity, and $Y$ is a microscopic quantity.
The notation $O(Z)$ will have its usual
meaning, i.e., $O(Z)=ZO(1)$, and $|O(1)|$ is bounded from above, but
it will be added that $|O(1)|<1/|\O|$. The $O(1)$ will be considered
dimensionless.

For a general spherically symmetric spacetime, the Lorentzian subspace
referred to below is its section at fixed angles, $4\pi r^2$ is the
area of the symmetry orbit passing through a given point, $v$ and $u$ are
the advanced and retarded times with past-directed gradients, labelling
the radial past and future light cones respectively. Assuming the
asymptotic flatness, $v$ is normalized at the past null infinity ($\mI$) as
\begin{equation}
(\nabla r,\nabla v)\scm =1\;.
\end{equation}
For one choice of $u$, $u=u^+$, a similar normalization holds at the
future null infinity ($\I$):
\begin{equation}
(\nabla r,\nabla u^+)\| =-1\;.
\end{equation}
Some other choices are considered below. The additive normalizations
of $v$ and $u$ are left arbitrary. The partial
derivatives $\partial_u$ and $\partial_v$ are defined as the derivatives
along the lines $\v$ and $\u$ respectively. The mapping on the 2-dimensional
Lorentzian subspace enables one to speak of points and lines instead
of 2-spheres and spherically symmetric hypersurfaces.

The curvature tensor of a spherically symmetric spacetime can be reduced
to two independent scalars. With $\triangle$ the D'Alembert operator
in the Lorentzian subspace, they can be introduced as
\begin{equation}
E=\frac{r}{2}\left(1-(\nabla r)^2\right)\;,\qquad
B=r\mathop{\triangle}r\;.
\end{equation}
The Riemann tensor (of the full 4-dimensional metric) is expressed
through these scalars in a differential manner, and their vanishing
in a domain is necessary and sufficient for the spacetime in this
domain to be flat. The function $E$ determines the ADM and Bondi
masses as, respectively, its limits at the spatial and null
infinities, and the (black) apparent horizon as the hypersurface
\begin{equation}
(\nabla r)^2=0\;,\qquad
(\nabla r,\nabla v)\ne0\;.
\end{equation}
For the consideration of the Bondi charges see section 6.

A spherically symmetric $T^{\mu\nu}$ has four algebraically independent
components with two differential constraints imposed on them by the
conservation equation. The constraints are solved explicitly by
expressing $T^{\mu\nu}$ through $E$ and $B$. For the components
of $T^{\mu\nu}$, the following basis is introduced:
\begin{equation}
A=4\pi r^2\left(T^{\mu\nu}\nabla_{\mu}u
\nabla_{\nu}u\right)\frac
{(\nabla r,\nabla v)^2}
{(\nabla u,\nabla v)^2}\;,
\end{equation}
\begin{equation}
D=4\pi r^2\left(T^{\mu\nu}\nabla_{\mu}v
\nabla_{\nu}v\right)\frac{1}
{(\nabla r,\nabla v)^2}\;,
\end{equation}
\begin{equation}
T_1 =4\pi r^2\left(T^{\mu\nu}\nabla_{\mu}u
\nabla_{\nu}v\right)\frac{2}
{(\nabla u,\nabla v)}\;,
\end{equation}
\begin{equation}
T_1 + T_2 =4\pi r^2 T^{\mu\nu}g_{\mu\nu}\;.
\end{equation}
Here $T_1$ is $4\pi r^2$ times the trace of $T^{\mu\nu}$ in
the Lorentzian subspace, $T_2$ is $4\pi r^2$ times the trace
of $T^{\mu\nu}$ in the complementary subspace, $A$ and $D$ govern
the speed of expansion of the radial light cones (eqs. (4.8)
and (4.81) below).

The expressions for $T^{\mu\nu}$ through $E$ and $B$ are
\begin{equation}
B=1-(\nabla r)^2 + T_1\;,
\end{equation}
\begin{eqnarray}
(\nabla r,\nabla v)\,\partial_v E&\!=\!&
A-\frac{1}{4}T_1\, (\nabla r)^2\;,\\
(\nabla r,\nabla u)\,\partial_u E&\!=\!&
\frac{1}{4}D\left( (\nabla r)^2\right)^2 -
\frac{1}{4}T_1\, (\nabla r)^2\;,
\end{eqnarray}
\begin{eqnarray}
T_2\, (\nabla r)^2 &\!=\!&
BD(\nabla r)^2 +
(\nabla r,\nabla v)\left(r\partial_v D\right)
(\nabla r)^2 +
(\nabla r,\nabla u)\left(r\partial_u T_1\right)
\nonumber\\
&\!=\!&
4AD+4\frac{(\nabla r,\nabla u)}
{(\nabla r)^2}\left(r\partial_u A\right)+
(\nabla r,\nabla v)\left(r\partial_v T_1\right)\;.
\end{eqnarray}
}

\newpage

{\renewcommand{\theequation}{2.\arabic{equation}}

\begin{center}
\section{\bf    Correspondence principle}
\end{center}

$$ $$

The correspondence principle for the collapse problem under
condition (1.2) can be formulated in terms of any congruence of
falling observers or falling light. It concerns the observables
of geometry which are, generally, scalar functions $I(x)$ of
a spacetime point $x$, and functionals of the geometry. The
correspondence principle is the assertion that, under certain
limitations to be discussed below, the values of observables
as measured in the units of $M$ by a given observer at a given
instant of his proper time differ from the classically predicted
values by $\O$. Both the proper time and the parameters identifying
the observer are supposed to be measured in the units of $M$.

Under the presently considered symmetry, it is convenient to choose
for the congruence of falling observers the family of radial past
light cones $\v$ The proper time of the observer is then replaced
by the affine time $\tau$ along $\v$ normalized at $\mI$ as
\begin{equation}
\left.\left(\frac{d\hphantom{\tau}}{d\tau}\,r
\Bigl|_{v=\mbox{\scriptsize const.}}\right)\right|_{\mI}=-1\;.
\end{equation}
With this normalization, the respective exact equation is of the form
\begin{equation}
\frac{d\hphantom{\tau}}{d\tau}\,r
\Bigl|_{v=\mbox{\scriptsize const.}}=-\rv\;.
\end{equation}
The observers' $v$ and $\tau$ specify the points $x$ of
the Lorentzian subspace. For functions $I(x)$ on this subspace
the assertion of the correspondence principle is
\begin{equation}
I(v,\tau)=I_{\mbox{\scriptsize class}}(v,\tau)+\O
\end{equation}
where $I_{\mbox{\scriptsize class}}(v,\tau)$ is the value 
of $I$ predicted for the given $v$ and $\tau$ by classical theory.

The limitations on the validity of eq. (2.3) concern both the
observers and the observables. One expects that, as the test
light ray $\v$ reaches the values of $r$ as small as $r=\O$, the
correspondence principle may cease being valid. However, this is not
the only limitation. Let us confine ourselves to the 
region $r>|\O|$, and let $v_0$ be some value of $v$ for which,
in this region, the correspondence principle is valid. It will be valid
as well for any value of $v$ differing from $v_0$ by a finite
multiple of $M$ but one cannot guarantee that it will remain
valid for $v$ as large as $v_0 + M/|\O|$ because, at the classical
limit, for being able to emit such a test light ray from $\mI$, one
should live for an infinitely long time. This aspect of the 
correspondence principle has been emphasized in ref. [4]. There may 
exist a critical value of $v$ defined with accuracy $O(M)$:
\begin{equation}
v_{\mbox{\scriptsize crit}}(\lambda)-v_0 =\frac{M}{|\O|}+O(M)
\end{equation}
such that, for $v>v_{\mbox{\scriptsize crit}}$, the correspondence
principle is no longer valid. Even at large $r$ but $v$ also 
large, the geometry will be nonclassical if the light cone $\v$ is
crossed by radiation.

The latter limitation on the validity of the correspondence principle
has its own limitation. A significant vacuum radiation can occur 
only at sufficiently late $u$, not earlier than the red shift
will become large:
\begin{equation}
\frac{du^+}{du^-}\gg 1\;.
\end{equation}
Here $u^-$ is the retarded time counted out by an early falling observer:
\begin{equation}
u^-=2\tau\Bigl|_{v=v_0}\;.
\end{equation}
In eq. (2.6), one can replace $v_0$ with any value $v_0 +O(M)$. This
will alter $du^+/du^-$ only by a factor of order 1.

To establish the bound in $u$ critical for the correspondence 
principle, consider an outgoing light ray $\u$ that crosses the
line $v=v_0$ at some $r>|\O|$. If, on its way 
to $v=v_{\mbox{\scriptsize crit}}$, this ray does not meet with
small $r$, one can use the classical geometry to calculate what
will be its $r$ at $v=v_{\mbox{\scriptsize crit}}$. Its classical
law of motion is
\begin{equation}
\makebox[14cm][l]{$\displaystyle \u\,\colon\qquad
\frac{r}{2M}+\ln\left|\frac{r}{2M}-1\right|
=\frac{v-v_0}{4M}+\left.\left(\frac{r}{2M}+\ln\left|
\frac{r}{2M}-1\right|\right)\right|_{v=v_0}\;.$}
\end{equation}
Hence one finds that if
\begin{equation}
\makebox[14cm][l]{$\displaystyle
\u\,\colon\qquad\qquad\qquad\qquad r\Bigl|_{v=v_0}-2M>|\O|\;, $}
\end{equation}
then $r$ along the ray $\u$ grows with $v$, so that eq. (2.7) can
be used indeed, and, at $v=v_{\mbox{\scriptsize crit}}$, this ray
comes to be already in the asymptotically flat region:
\begin{equation}
\makebox[14cm][l]{$\displaystyle
\u\,\colon\qquad\qquad\qquad\qquad r\Bigl|_{v=v_{\mbox{\scriptsize crit}}}
=\frac{M}{|\O|}\;.$}
\end{equation}
At $v<v_{\mbox{\scriptsize crit}}$ it passes across the region
of classical geometry, and
at $v>v_{\mbox{\scriptsize crit}}$ it passes across the
asymptotically flat region where the geometry may differ
from the classical one only if the red shift is already large.
The red-shift factor in eq. (2.5) may be written as a product
of two
\begin{equation}
\frac{du^+}{du^-}=\frac{du^+}{du^*}\frac{du^*}{du^-}
\end{equation}
with
\begin{equation}
u^* =2\tau\Bigl|_{v=v_{\mbox{\scriptsize crit}}}\;.
\end{equation}
The first factor in this product involves only the asymptotically
flat region. The red shift cannot accumulate in this region
since the curvature in it is everywhere small:
\begin{equation}
\frac{du^+}{du^*}=1+\O\;.
\end{equation}
The second factor involves only the region of classical geometry
and, therefore, can be calculated:
\begin{equation}
\makebox[14cm][l]{$\displaystyle \u\,\colon\qquad\qquad\qquad\qquad
\frac{du^*}{du^-}=\frac{\textstyle 1-\left.\left(2M/r\right)
\right|_{v=v_{\mbox{\scriptsize crit}}}}
{\textstyle 1-\left.\left(2M/r\right)
\right|_{v=v_0}}\;.$}
\end{equation}
Under condition (2.8) one obtains
\begin{equation}
\frac{du^*}{du^-}<\frac{1}{|\O|}\;.
\end{equation}
Thus, for $u$ satisfying condition (2.8), the red shift is still
moderate, and, therefore, the line $\u$ with this value 
of $u$ lies entirely in the region of classical geometry.

Denote as $u_0$ the value of $u$ for which
\begin{equation}
\makebox[14cm][l]{$\displaystyle u=u_0\,\colon\qquad\qquad\qquad\qquad
r\Bigl|_{v=v_0}=2M(1+\O_0)$}
\end{equation}
with some chosen $\O_0$. We found two (overlapping) regions in the
vacuum in which the correspondence principle is valid, CL.I and CL.II:
\begin{equation}
\makebox[14cm][l]{$\displaystyle \mbox{CL.I}\,\colon\qquad
\qquad\qquad\qquad\qquad r>|\O|\;,\quad v<v_{\mbox{\scriptsize crit}}\;,$}
\end{equation}
\begin{equation}
\makebox[14cm][l]{$\displaystyle \mbox{CL.II}\,\colon\qquad
\qquad\qquad\qquad\qquad\quad u^-<u^-_0 -|\O|\;.$}
\end{equation}
Their union will be denoted as CL and called classical region.
Note that the classical region is {\it causally complete} in a
sense that it contains all of its causal past (in the vacuum).
For CL.I, this follows from the fact that, by the classical
geometry, the line $r=|\O|$ in the vacuum is spacelike.
It should be emphasized that for local or retarded equations with 
data in the past, as the expectation-value equations are in any
approximation, the causal completeness of the domain of validity
of the approximation is a necessary condition for obtaining the 
solution.

As mentioned above, the limitations on the validity of the 
correspondence principle concern also the observables $I(x)$.
If the point $x$ is in the classical region but the dependence
of $I(x)$ on the geometry is nonlocal, then $I(x)$ may involve
not only the classical region. The correspondence principle
is valid deliberately only for local and retarded observables,
i.e., the functions $I(x)$ that depend on the geometry only
at~$x$ and in the past of~$x$. Important examples are the
scalars $\rv$ and $(\nabla r,\nabla u^+)$.
The former is a retarded observable whereas the latter is an
advanced one. Therefore, even in the classical region,
$(\nabla r,\nabla u^+)$ may differ drastically 
from its classical value (see below). On the other hand,
by the correspondence principle,
\begin{equation}
\makebox[14cm][l]{$\displaystyle \mbox{CL}\,\colon
\qquad\qquad\qquad\qquad\qquad \rv=1+\O\;.$}
\end{equation}
Eq. (2.2) then integrates to give
\begin{equation}
\makebox[14cm][l]{$\displaystyle \mbox{CL}\,\colon\qquad
\qquad\qquad\qquad\qquad\qquad \tau=-r+f(v)\;.$}
\end{equation}
Using this relation in eq. (2.3) one obtains the final formulation
of the correspondence principle. This is the assertion
\begin{equation}
I(v,r)=I_{\mbox{\scriptsize class}}(v,r)+\O
\end{equation}
valid for all local and retarded observables $I$ in the union
of regions (2.16) and (2.17).

By the correspondence principle, in the classical region, in an 
appropriate vector basis, $\T$ should be small. For being
appropriate, the vector basis must satisfy two 
requirements: (i) it should be nonsingular in the classical 
geometry, and (ii) the basis vectors may depend on the
metric only in a local or retarded manner. Then, since $\T$ is
retarded, its projections on the basis vectors will be retarded
observables. The following vector basis in the Lorentzian subspace
meets these requirements:
\begin{equation}
\mathop{\nabla}\nolimits_{\mu}v\;,\qquad
(\nabla v,\nabla u)^{-1}
\mathop{\nabla}\nolimits_{\mu}u\;.
\end{equation}
Note that the second vector is independent of the normalization 
of $u$. The respective coordinate basis is the one in which
the spacetime points are labelled with the values 
of $v$ and $\tau\bigl|_{v=\mbox{\scriptsize const.}}$. This is 
the labelling accomplished
by the presently chosen congruence of falling observers. The
vectors (2.21) are retarded. Dividing the first of them and
multiplying the second by $\rv$, one can build a purely local basis
satisfying the same requirements:
\begin{equation}
{\rv}^{-1}\mathop{\nabla}\nolimits_{\mu}v\;,\qquad
\rv (\nabla v,\nabla u)^{-1}
\mathop{\nabla}\nolimits_{\mu}u\equiv (\partial_u r)
\mathop{\nabla}\nolimits_{\mu}u\;.
\end{equation}
This is the vector basis used in eqs. (1.8)-(1.11). Therefore,
by the correspondence principle,
\begin{equation}
\makebox[14cm][l]{$\displaystyle \mbox{CL}\,\colon
\qquad\qquad\qquad\qquad A,\; D,\; T_1 ,\; T_2 =\O\;,$}
\end{equation}
and for the basic curvature scalars in eq. (1.6) one has
\begin{equation}
\makebox[14cm][l]{$\displaystyle \mbox{CL}\,\colon
\qquad\qquad\qquad E=M(1+\O)\;,\qquad 
B=\frac{2M}{r}+\O\;.$}
\end{equation}

The projections of covariant derivatives of $\T$ on the 
vectors (2.22) also are retarded observables and, by the
correspondence principle, should also be $\O$ in the classical
region. By the dimension of the coupling constant, the
differential order of the expectation-value spacetime
is at least ${\rm C}^4$, as distinct from the classical ${\rm C}^2$.
In particular, the local terms of $\T$ contain at least the
second-order derivatives of the curvature. Therefore, one disposes
of conditions on the first and second derivatives of $\T$. These
conditions are obtained, as one can check, by acting with linear
and quadratic combinations of the operators
\begin{equation}
M\partial_v\qquad \mbox{and}\qquad M(\partial_u r)^{-1}\partial_u
\end{equation}
on the scalars in eq. (2.23), and, in CL, equating the results
to $\O$.
}

\newpage

{\renewcommand{\theequation}{3.\arabic{equation}}

\begin{center}
\section{\bf    Correspondence principle (continued)}
\end{center}

$$ $$

Consider now the spacetime region complementary to the classical
one. It is hardly possible that, at $r=\O$, the curvature in the
units of $M$ does not become as large as $1/\O$. Then, at $r=\O$,
not only classical theory is invalid. In a region where the curvature 
is $1/\O$, the indefinite local terms of $\T$ cease being small.
This invalidates semiclassical theory as well. However, one does not
expect that the curvature becomes that large at $r>|\O|$ including
at $v>v_{\mbox{\scriptsize crit}}$. This is potentially the region 
of validity of semiclassical theory. The ultraviolet problem bears
apparently no relation to it. The reserves "potentially" and
"apparently" are made because the region $r>|\O|$ may be not
causally complete. If there are small $r$ in its causal past in
the vacuum, one will not be able to use semiclassical theory
in this region. The region $r=\O$ together with its causal future
ought to be, rightfully, called ultraviolet region and excluded
from the consideration. Semiclassical region (denoted below 
as SCL) is, by definition, the region of validity of semiclassical
theory.

To establish the bounds set by causality, consider again an
outgoing light ray $\u$ that, at $v=v_0$, has $r>|\O|$. Before
this ray reaches $v=v_{\mbox{\scriptsize crit}}$ or meets with
small $r$, the law of its motion is the one in eq. (2.7). One
finds that if
\begin{equation}
\makebox[14cm][l]{$\displaystyle \u\,\colon\qquad\qquad\qquad\qquad
r\Bigl|_{v=v_0}-2M<-|\O|\;,$}
\end{equation}
then $r$ along this ray decreases with $v$ and reaches the 
value $|\O|$ at
\begin{equation}
\makebox[14cm][l]{$\displaystyle\u\,\colon\qquad\qquad\qquad\qquad
v\Bigl|_{r=|\O|}=v_0 + |O(M)|\;.$}
\end{equation}
The future of such a light ray is not predictable either by classical
or by semiclassical theory. Its future beyond the value of $v$ in
eq. (3.2) is in the ultraviolet region, and its past is in the
classical region.

Eqs. (2.8) and (3.1) leave for the semiclassical region only the
values of $u$ for which
\begin{equation}
r\Bigl|_{v=v_0}=2M(1\pm |\O|)\;,
\end{equation}
i.e., only the interval
\begin{equation}
\makebox[14cm][l]{$\displaystyle\mbox{SCL}\,\colon
\qquad\qquad\qquad\qquad\qquad
u^-=u^-_0\pm |\O|\;,$}
\end{equation}
\begin{equation}
\makebox[14cm][l]{$\displaystyle\hphantom{\mbox{SCL}\,\colon
\qquad\qquad\qquad\qquad\qquad}\,
v>v_{\mbox{\scriptsize crit}}$}
\end{equation}
with some range of $\O$ in eq. (3.4). It will be emphasized
that, in particular, the calculation of the radiation 
at $\I$, whatever semiclassical technique is used, the effective
action or WKB, is valid only in the interval (3.4) of $u$.
Along $v=v_0$, this interval is a microscopic neighbourhood
of $r=2M$. An early falling observer crosses it without noticing
because the whole of this interval is within the quantum 
uncertainty of measuring of his proper time. However, because of
a large red shift, an interval microscopic in $u^-$ may be
macroscopic and even infinite in $u^+$. The later falls the 
observer, the longer is, for him, this interval. For the
incoming light signal with $v=v_{\mbox{\scriptsize crit}}$, the
length of this interval in $\tau$ and $r$ may already equal
units of $M$, and, possibly, this interval covers the whole
of the future of $\I$.

In the semiclassical region, $\T$ and its derivatives may be 
not small any more but, as a matter of principle, their projections
on the basis vectors (2.22) remain bounded. Specifically,
\begin{equation}
\makebox[14cm][l]{$\displaystyle\mbox{SCL}\,\colon
\qquad\qquad\qquad\qquad\qquad A,\;D,\;T_1 ,\;T_2 =O(1)\;,$}
\end{equation}
\begin{equation}
\makebox[14cm][l]{$\displaystyle\hphantom{\mbox{SCL}\,\colon
\qquad\qquad\qquad\qquad\qquad}
M(\partial_u r)^{-1}\partial_u A=O(1)\;,$}
\end{equation}
\begin{equation}
\makebox[14cm][l]{$\displaystyle\hphantom{\mbox{SCL}\,\colon
\qquad\qquad\qquad\qquad\qquad}
M(\partial_u r)^{-1}\partial_u D=O(1)\;,$}
\end{equation}
\begin{equation}
\makebox[14cm][l]{$\displaystyle\hphantom{\mbox{SCL}\,\colon
\qquad\qquad\qquad\qquad\qquad}
M(\partial_u r)^{-1}\partial_u B=O(1)\;,$}
\end{equation}
\begin{equation}
\makebox[14cm][l]{$\displaystyle\hphantom{\mbox{SCL}\,\colon
\qquad\qquad\qquad\qquad\qquad}\,
M^2\left((\partial_u r)^{-1}\partial_u\right)
   \left((\partial_u r)^{-1}\partial_u\right)A=O(1)\;,$}
\end{equation}
\begin{equation}
\makebox[14cm][l]{$\displaystyle\hphantom{\mbox{SCL}\,\colon
\qquad\qquad\qquad\qquad\qquad}\,
M^2(\partial_u r)^{-1}\partial_u\partial_v B=O(1)\;.$}
\end{equation}
Here use is made of the operators (2.25), and the conditions
for $B$ are obtained from the conditions for $T_1$. In addition,
with a certain reserve one may use that
\begin{equation}
\makebox[14cm][l]{$\displaystyle\mbox{SCL}\,\colon
\qquad\qquad\qquad\qquad
|\O|<\;\rv\;,\;\frac{E}{M}\;,\;\frac{rB}{2E}\;<\frac{1}{|\O|}$}
\end{equation}
because, in the classical region, these quantities equal 1. In
the semiclassical region, they may eventually turn into zero
or infinity but, before that, there will be an evolution.
Obtaining this evolution is one's goal.

By the correspondence principle, the apparent horizon (AH) enters
the vacuum region CL.I with
\begin{equation}
\frac{d\rH}{dv_{\hphantom{\mbox{\tiny AH}}}}\ge 0
\end{equation}
and, in this region, has $\rH=2M(1+\O)$ at all $v$. Therefore,
the AH initially gets into the interval (3.4) of $u$. As it evolves,
it cannot go out of this interval to smaller $u$ for, otherwise,
it will get to the region CL.II at variance with the correspondence
principle. It can go out of this interval to greater $u$ but only
when having already $\rH=\O$. Indeed, a greater-$u$ outgoing
ray is the one in eqs. (3.1), (3.2). Along it, $r$ decreases
monotonically down to $r=\O$. If, instead of getting to a 
singularity, this ray crosses the AH, then only at $r=\O$. It
follows that the semiclassical region (3.4) covers the whole
of the evolution of the AH from $\rH=2M$ to $\rH=\O$.
The latter value may be not reached. Then the AH stays in the 
semiclassical region always.

Suppose that eq. (3.13) holds throughout the interval (3.4).
Then one of the lines $\u$ in this interval is an event horizon
hiding a black hole of mass greater than or equal to the ADM mass.
This is at variance with the radiation of positive energy. 
Therefore, there should be a point of the AH at which the
derivative in eq. (3.13) changes the sign. At this point, the
AH is exactly null and tangent to one of the lines $\u$ in the
interval (3.4). It is natural to assume that this point is 
in CL.I where the deviation of the AH from a null line is within 
the quantum uncertainty. One can then choose the $\O_0$ in
eq. (2.15) and shift $v_0$ by $O(M)$ so that this point 
be $(u_0 , v_0)$. To the past from this point in the advanced
time, the AH is spacelike, and to the future timelike. The
outgoing rays with $u<u_0$ never cross it, and the ones
with $u>u_0$ cross it twice. The first crossing occurs in the
support of $\t$ or in the Planckian neighbourhood of this
support [5]. The tangency point $(u_0 , v_0)$ is in the band
of quantum uncertainty around the support of $\t$. Therefore, when
considering the vacuum region, one may confine oneself 
to $v\ge v_0$. This limitation is implied below. Respectively,
unless the context assumes otherwise, the discussion of the AH 
below refers to its second, in the order in which the light 
rays $\u$ cross it, branch. The behaviour of the first branch
in the vacuum is subject to a different physics: creation of
the virtual pairs as opposed to the real ones. The effect 
of the local vacuum polarization is considered in ref. [5].
The AH is shown in Fig. 1 for not very late $u$. The point 0
in Fig. 1 is $(u_0 , v_0)$.

Even after crossing the AH the second time, the light 
rays $\u$ might not go out to $\I$ but, if the second branch
of the AH is caused by radiation, they do. Then the 
chart $u^+$ extends to the AH. Consequently,
\begin{equation}
(\nabla v,\nabla u^+)
\Bigl|_{\mbox{\tiny AH}}\,\ne 0,\;\pm\infty\;,
\end{equation}
\begin{equation}
(\nabla r,\nabla u^+)
\Bigl|_{\mbox{\tiny AH}}\, =0
\hphantom{,\;\pm\infty\;,} 
\end{equation}
including at the point 0 where the geometry is "most classical".
Eq. (3.15) is at no variance with the fact that the classical
value of the {\it advanced} 
observable $(\nabla r,\nabla u^+)$ is $-1$.

The observable $-(\nabla v,\nabla u^+)/2$ is the
red-shift factor that the outgoing light signal with the current
value of $u$ accumulates while passing from the current value
of $v$ to $\I$:
\begin{equation}
-\frac{1}{2}\,(\nabla v,\nabla u^+)=
\frac{du^+_{\hphantom{\mbox{\scriptsize current }v}}}
{du^{\hphantom{+}}_{\mbox{\scriptsize current }v}}\;,
\end{equation}
$$
u_{\mbox{\scriptsize current }v}=
2\tau\Bigl|_{\mbox{\scriptsize current }v}\;.
$$
The full red-shift factor is
\begin{equation}
-\frac{1}{2}\,(\nabla v,\nabla u^+)\Bigl|_{v=v_0}
=\frac{du^+}{du^-}\;.
\end{equation}
In the classical geometry, the observable (3.16) turns into
infinity at the AH because the red shift becomes infinite.
The advanced observable
\begin{equation}
\frac{\partial r^{\hphantom{+}}}{\partial u^+}=\frac{1}{2}\,
\frac{\rr}{(\nabla r,\nabla u^+)}=
\frac{\rv}{(\nabla v,\nabla u^+)}
\end{equation}
differs from $(\nabla v,\nabla u^+)^{-1}$ only
by the coefficient $\rv$ of order 1 (in CL, just 1). Therefore,
in the classical geometry, it vanishes at the AH. In the
expectation-value geometry, the observable (3.16) remains finite
at the AH by virtue of eq. (3.14). As a consequence, the observable
(3.18) does not vanish:
\begin{equation}
\left.\frac{\partial r^{\hphantom{+}}}{\partial u^+}
\right|_{\mbox{\tiny AH}}=
\frac{d\hphantom{u^+}}{du^+}\, \rH=
2\,\frac{d\hphantom{u^+}}{du^+}\, \EH\ne 0
\end{equation}
and determines the law by which $r$ and $E$ vary along the AH.
It does not vanish, in particular, at the point 0 where the AH
is null. At this point one has
$$
(\partial_v r)\Bigl|_0=0\;,\quad
(\partial^2_{vv}r)\Bigl|_0=0\;,\quad
(\partial_v E)\Bigl|_0=0\;,
$$
\begin{equation}
\left.\frac{d\uH^+}{dv^{\hphantom{+}}_{\hphantom{\mbox{\tiny AH}}}}
\right|_0=0\;,\quad
\left.\frac{d\rH}{dv_{\hphantom{\mbox{\tiny AH}}}}
\right|_0=0\;,\quad
\left.\frac{d\rH^{\hphantom{+}}}{du^+_{\hphantom{\mbox{\tiny AH}}}}
\right|_0\ne 0\;.
\end{equation}

The semiclassical region is in the future domain of dependence
of the classical region. With the classical region 
included, it is causally complete. Therefore, the correspondence
principle plays the role of the initial condition for the
expectation-value equations in the semiclassical region.
Now I go over to the question where these equations will come
from if $\T$ is not to be calculated except at $\I$. Below,
only the semiclassical region is considered.
}

\newpage

{\renewcommand{\theequation}{4.\arabic{equation}}

\begin{center}
\section{\bf    Solution in the region of strong field}
\end{center}

$$ $$

The key point is that, in the region where the outgoing light
signals acquire a large red shift:
\begin{equation}
\frac{\partial r^{\hphantom{+}}}{\partial u^+}=\O\;,
\end{equation}
i.e., in the region of strong gravitational field, one does not 
need field equations. The conditions of boundedness of the curvature
(3.6)-(3.11) take the place of the field equations in this region.
First note that, since, by eq. (3.12), $\rv >0$ and $B>0$, one has
\begin{equation}
\dr <0\;,\qquad\quad \partial_v\left|\dr\right|>0\;.
\end{equation}
Furthermore, for $\rr >|\O|$ one has
\begin{equation}
\makebox[14cm][l]{$\displaystyle \rr >|\O|\,\colon
\qquad\qquad\qquad\qquad (\nabla r,\nabla u^+)=O(1)$}
\end{equation}
and hence
\begin{equation}
\makebox[14cm][l]{$\displaystyle \rr >|\O|\,\colon
\qquad\qquad\qquad\qquad \left|
\frac{\partial r^{\hphantom{+}}}{\partial u^+}\right| >|\O|\;.$}
\end{equation}
Eq. (4.3) is a consequence of the boundedness condition $A=O(1)$,
and its proof repeats with an obvious modification the derivation 
of eq. (6.26) below. Eq. (4.4) then follows from the identity (3.18).
Therefore, condition (4.1) implies $\rr \le |\O|$, and it is natural 
to assume that the apparent horizon $\rr =0$ is in the region (4.1):
\begin{equation}
\left.\frac{\partial r^{\hphantom{+}}}{\partial u^+}
\right|_{\mbox{\tiny AH}}=\O\;.
\end{equation}
Then, outside the AH, i.e., at $\rr >0$, condition (4.1) holds as 
long as $\rr =|\O|$, and, inside the AH, it holds as far as the
rays $\u$ extend by virtue of eqs. (4.2) and (4.5). It follows that,
in the chart $u^+$, the region (4.1) covers the union
\begin{equation}
\Bigl(\rr =|\O|\Bigr)\;\bigcup\; \Bigl(\rr <0\Bigr)\;.
\end{equation}
Below it will be shown that, at $\rr <0$, the chart $u^+$ does not
extend beyond ${\rr =-|\O|}$.

In the region (4.6), one can replace $M$ in eq. (3.9) with $r$
replacing first $M$ with $E$ on the basis of condition (3.12)
and next $E$ with $r$ on the basis of eq. (4.6). This gives the
equation
\begin{equation}
r\partial_u B=O(1)\dr\;,
\end{equation}
and the following equations hold identically:
\begin{equation}
r\partial_u\ln \rv =-D\dr\;,
\end{equation}
\begin{equation}
r\partial_u\ln r=\dr\;,
\end{equation}
\begin{equation}
r\partial_v\ln\left| \dr\right|=\frac{B}{2\rv}\;.
\end{equation}
With any choice of $u$ for which $\partial r/\partial u =\O$, not
necessarily $u=u^+$, eqs. (4.7)-(4.10) close to lowest order in
$\partial r/\partial u$. This is a consequence of two boundedness 
conditions: eq. (3.9) and $D=O(1)$, which thus play the role
of the field equations.

The solution of eqs. (4.7)-(4.10) is given below, and its derivation
will be found in section 7, but one obvious question should be
answered right away. Eqs. (4.7)-(4.10) are invariant with respect
to the choice of $u$, and there are choices for which
$\partial r/\partial u$ is not small. Then what is the use of these
equations? The answer is that the initial data to these equations
break the invariance. An equation like (4.7) is usable only if
its right-hand side contributes negligibly to the solution
throughout the region (4.1). One possibility for that is to have
$$
K>|\O|
$$
where
$$
-K=\partial_u\ln \left|\dr\right|\;.
$$
This will prove to be the case below. However, $K$ is not invariant
with respect to the choice of $u$. It is necessary that the condition
$K>|\O|$ hold with the same choice of $u$ with which the condition
$\partial r/\partial u =\O$ holds, i.e., $u=u^+$. This will be secured
by the properties of the data functions.

Since the AH is in the region (4.1), the initial data to eqs. 
(4.7)-(4.10) can be taken at the AH. Four data functions are needed:
\begin{equation}
\EH\;,\;\BH\;,\;\rv\Bigl|_{\mbox{\tiny AH}}\;,\;\vH (u)
\end{equation}
where $v=\vH (u)$ or, conversely, $u=\uH (v)$ is the equation of
the second branch of the AH. The data for $r$ and 
$\partial r/\partial u$ are expressed through $\EH$, and the
following notation is introduced:
\begin{equation}
\alpha =\rv\Bigl|_{\mbox{\tiny AH}}\;,\qquad 
\beta =-\left.\dr\right|_{\mbox{\tiny AH}}
=-2\,\frac{d\EH}{du_{\hphantom{\mbox{\tiny AH}}}}\;,
\end{equation}
\begin{equation}
\kappa =-\frac{d\ln\beta}{du\hphantom{\ln}}+
\frac{\BH}{4\alpha\EH}\,\frac{d\vH}{du_{\hphantom{\mbox{\tiny AH}}}}\;.
\end{equation}
Note that $\beta$ and $\kappa$ are not invariant with respect
to the choice of $u$. Throughout the present section, $u$ is $u^+$ or 
differs from $u^+$ by a finite transformation. By eqs. (3.12) and
(4.5),
\begin{equation}
|\O|<\;\alpha\;,\;\EH\;,\;\BH\;<\frac{1}{|\O|}\;,\qquad
\beta =|\O|\;,
\end{equation}
and on the same grounds one may use that
\begin{equation}
\frac{d\uH}{dv_{\hphantom{\mbox{\tiny AH}}}}<\frac{1}{|\O|}\;.
\end{equation}
It will additionally be assumed and next confirmed that
\begin{equation}
\frac{d\BH}{dv_{\hphantom{\mbox{\tiny AH}}}}=\O
\end{equation}
and
\begin{equation}
|\O|<\;\kappa\;<\frac{1}{|\O|}\;,\qquad
\frac{d\hphantom{u}}{du}\,\frac{1}{\kappa}=\O\;.
\end{equation}
Conditions (4.16) and (4.17) on the data functions will be derived in
section 7 when sewing together the solutions in the regions
of strong and weak field. They are conditions of the existence 
of the global solution. The data functions taken at the second 
branch of the AH at the point with a given value of $u$ will
be denoted as $\EH (u)$, $\alpha (u)$, etc. The same functions 
taken at the point of the AH with a given value of $v$ will be 
denoted as $\EH (v)$, $\alpha (v)$, etc. The data functions
satisfy the identity following from eq. (4.13)
\begin{equation}
\beta (u){\rm e}^{\Gamma_1}=\beta (v){\rm e}^{\Gamma_2}\;,
\qquad v\ge v_0
\end{equation}
where $\Gamma_1$ and $\Gamma_2$ are the integrals along the
second branch of the AH
\begin{equation}
\Gamma_1 =\int\limits^v_{\vH (u)}dv\,\frac{\BH}{4\alpha\EH}\;,
\qquad\Gamma_2 =\int\limits_u^{\uH (v)}du\,\kappa\;.
\end{equation}
The values of the data functions at the point 0 will be denoted
as $\alpha_0$, $\beta_0$, $\kappa_0$, etc.

With the data as above, the solution in the region (4.1) for
$v\ge v_0$ is
\begin{equation}
(1+\O)\dr=-\beta (u){\rm e}^{\Gamma_1}=
-\beta (v){\rm e}^{\Gamma_2}\;,
\end{equation}
\begin{equation}
(1+\O)\rv =\alpha (v)\;,
\end{equation}
\begin{equation}
(1+\O)\uv^{-1}=-\frac{\beta (v)}{\alpha (v)}{\rm e}^{\Gamma_2}\;,
\end{equation}
\begin{equation}
r=2\EH (v)+\beta (v)\left(\frac{1}{\kappa (u)}{\rm e}^{\Gamma_2}
-\frac{1}{\kappa (v)}\right)(1+\O)\;.
\end{equation}
The latter two equations give the metric in the null coordinates.
One obtains
\begin{equation}
(1+\O)\rr =2\beta (v)\frac{\BH (v)}{4\EH (v)}
\left(\frac{1}{\kappa (u)}{\rm e}^{\Gamma_2}
-\frac{1}{\kappa (v)}\right)\;,
\end{equation}
\begin{equation}
(1+\O)\ru =\frac{\BH (v)}{4\EH (v)}
\left(\frac{1}{\kappa (v)}{\rm e}^{-\Gamma_2}
-\frac{1}{\kappa (u)}\right)\;,
\end{equation}
\begin{equation}
E=\EH (v)+ \frac{1}{2}\beta (v)\Bigl(1-\BH (v)\Bigr)
\left(\frac{1}{\kappa (u)}{\rm e}^{\Gamma_2}
-\frac{1}{\kappa (v)}\right)(1+\O)\;,
\end{equation}
\begin{equation}
B=\BH (v)+\O\;,
\end{equation}
\begin{equation}
T_1 =\BH (v) -1+\O\;,
\end{equation}
\begin{eqnarray}
A&\!=\!&\alpha (v)\BH (v)
\frac{d\EH (v)}{dv\hphantom{\EH )(}}(1+\O)\nonumber\\
&\!=\!&-\frac{1}{2}\beta (v)\alpha (v)
\frac{d\uH (v)}{dv\hphantom{\uH )(}}\BH (v)(1+\O)\;.
\end{eqnarray}
Only the curvatures $D$ and $T_2$ are not obtained because
they are contained in the approximation of higher order in
$\partial r/\partial u$. In eqs. (4.20)-(4.29), the left-hand 
sides are functions of the observation point $(u,v)$, and
the coordinates of the observation point appear as the
arguments of the data functions on the right-hand sides.
For details of the derivation see section 7.

Under the restriction $v\ge v_0$ assumed in the solution above,
the exterior and interior of the AH are respectively the regions
\begin{equation}
\makebox[14cm][l]{$\displaystyle \rr >0\,\colon\qquad
v>\vH (u)\;,\quad u<\uH (v)\;,\quad \Gamma_1>0\;,\quad \Gamma_2>0\;,$}
\end{equation}
\begin{equation}
\makebox[14cm][l]{$\displaystyle \rr <0\,\colon\qquad
v<\vH (u)\;,\quad u>\uH (v)\;,\quad \Gamma_1<0\;,\quad \Gamma_2<0\;.$}
\end{equation}
The solution has two immediate consequences. First, from eq. (4.25)
one infers that the condition (4.3) extends to the whole of the
exterior of the AH at $v\ge v_0$:
\begin{equation}
\makebox[14cm][l]{$\displaystyle \rr >0\,\colon
\qquad\qquad\qquad\qquad \ru =O(1)\;.$}
\end{equation}
Second, for the interior of the AH, eq. (4.24) yields the bound
\begin{equation}
\rr >-2\beta (v)\frac{\BH (v)}{4\EH (v)\kappa (v)}(1+\O)
\end{equation}
to be discussed below.

Another important consequence of the solution is that the
curvature $A$ proves to be $\O$. That $A$ at the AH is $\O$
follows directly from the exact equations (1.12)-(1.14).
Eqs. (1.12) and (1.14) imply
\begin{equation}
(\partial_u E)\Bigl|_{\mbox{\tiny AH}}=
(1-\BH)\frac{d\EH}{du_{\hphantom{\mbox{\tiny AH}}}}\;,\qquad
(\partial_v E)\Bigl|_{\mbox{\tiny AH}}=
\BH\frac{d\EH}{dv_{\hphantom{\mbox{\tiny AH}}}}\;,
\end{equation}
and hence by eq. (1.13)
\begin{equation}
A\Bigl|_{\mbox{\tiny AH}}=\alpha\BH
\frac{d\EH}{dv_{\hphantom{\mbox{\tiny AH}}}}\;.
\end{equation}
Eq. (4.29) adds to this fact that $A$ is constant along the lines
$\v$ and, therefore, is small throughout the region (4.1).
The smallness of $A$ and the boundedness of the second derivative
of $A$ (eq. (3.10)) imply the smallness of the first derivative
of $A$. The respective bound is obtained in the Appendix:
\begin{equation}
M(\partial_u r)^{-1}\partial_u A=O(1)|A|^{1/2}=O(\beta^{1/2}(v))\;.
\end{equation}

The differential equations along the lines $\v$, whose solutions
are obtained above, have counterparts along the lines $\u$ For
the function $r$, this is the identity
\begin{equation}
r\partial_v\ln r =\frac{\rr}{2\rv}\;,
\end{equation}
and, for the functions $E$ and $B$, these are eq. (1.13)
and the second form of eq. (1.15) with $T_1$ expressed through $B$.
Note the distinction: the right-hand side of eq. (4.37) is
$O(\rr)$ whereas the right-hand side of the analogous equation
(4.9) is $O(\partial r/\partial u)$. As explained below, this
distinction is essential but, outside the AH where eq. (4.32)
holds, there is no distinction:
\begin{equation}
\makebox[14cm][l]{$\displaystyle \rr >0\,\colon
\qquad\qquad\qquad\qquad
\rr =O\left(\dr\right)\;.$}
\end{equation}
Furthermore, outside the AH
\begin{equation}
\makebox[14cm][l]{$\displaystyle \rr >0\,\colon
\qquad\qquad\qquad\qquad
\beta (v)<\left|\dr\right|$}
\end{equation}
as follows from eq. (4.20). Therefore,
\begin{equation}
\makebox[14cm][l]{$\displaystyle \rr >0\,\colon\qquad\qquad
\beta (v)=O\left(\dr\right)\;,\qquad
\beta^{1/2}(v)=O\left(-\dr\right)^{1/2}\;,$}
\end{equation}
and eqs. (1.13) and (1.15) take the form
\begin{equation}
\makebox[14cm][l]{$\displaystyle \rr >0\,\colon\qquad\qquad
\partial_v E =O\left(\dr\right)\;,\qquad
r\partial_v B=O\left(-\dr\right)^{1/2}\;.$}
\end{equation}
Here use is made of eqs. (4.29) and (4.36). The differential 
equations (4.37)-(4.38) and (4.41) can be integrated along
the lines $\u$ with the aid of eqs. (4.10) and (3.12), e.g.,
\begin{equation}
\int\limits^v_{\vH (u)}\frac{dv}{r}\,
\left.O\left(-\dr\right)^{1/2}\right|_{u={\mbox{\scriptsize const.}}}
=O\left(-\dr\right)^{1/2}-O\left(\beta^{1/2}(u)\right)\;,
\end{equation}
and their solutions in the region of strong field are analogous
to the solutions obtained above by integrating the equations along
the lines $\v$:
\begin{equation}
\makebox[14cm][l]{$\displaystyle \rr >0\,\colon\qquad\qquad\qquad\qquad
r=2\EH (u)(1+\O)\;,$}
\end{equation}
\begin{equation}
\makebox[14cm][l]{$\displaystyle\hphantom{\rr >0\,\colon
\qquad\qquad\qquad\qquad}E=\EH (u)(1+\O)\;,$}
\end{equation}
\begin{equation}
\makebox[14cm][l]{$\displaystyle\hphantom{\rr >0\,\colon
\qquad\qquad\qquad\qquad}B=\BH (u)+\O\;.$}
\end{equation}

Then consider any point $(u,v)$ that belongs to the region (4.1) 
and is located outside the AH. In Fig. 1, this is point 1, and it
defines points 2 and 3. It follows from eqs. (4.23)-(4.27) and
(4.43)-(4.45) that, up to microscopic variations, the functions
$r$, $E$, and $B$ are constant in the triangle 123. As a consequence,
with the same accuracy, the data functions $\EH$ and $\BH$ are
constant in the sector 23 of the AH. Specifically, for the
points 2 and 3 one obtains the relations
\begin{equation}
\EH (v)=\EH (u)(1+\O)\;,\qquad\BH (v)=\BH (u)+\O
\end{equation}
valid if $u$ and $v$ satisfy the conditions
\begin{equation}
\left.\dr\right|_{\mbox{\scriptsize point }(u,v)}=\O\;,\qquad
\rr\Bigl|_{\mbox{\scriptsize point }(u,v)}>0\;.
\end{equation}
For the exterior of the AH, this property of the data functions
can be used in the solution (4.20)-(4.29).

The distinction between the equations along $\v$ and $\u$ is
essential inside the AH because, there, $\Gamma_2$ is negative
and may become as large as $-1/|\O|$. This makes condition (4.32)
invalid at $\rr <0$: inside the AH, the function $\ru$ may grow
up to $1/|\O|$. Then eqs. (4.38)-(4.40) will no longer be valid,
and solutions (4.43)-(4.45) will not apply. It is entirely owing
to this fact that $E$ and $B$ can undergo macroscopic variations
along the AH. Indeed, consider any line $\u$ crossing the AH. In
Fig. 1, it is shown as passing through the points 4 and 5. Since
\begin{equation}
\partial_v \Gamma_2 =\frac{d\uH (v)}{dv_{\hphantom{\mbox{\tiny AH}}}
\hphantom{(v)}}\kappa (v)\ge 0\;,
\end{equation}
$\Gamma_2$ decreases along this line from point 4 where it is zero
to point 5 where it has a minimum. If, at this minimum, $\Gamma_2$
is $O(1)$, the solutions (4.43)-(4.45) apply, and one has
$E_4=E_5(1+\O)$, $B_4=B_5+\O$. On the other hand, by eqs. (4.26)
and (4.27), $E_5=E_0(1+\O)$, $B_5=B_0+\O$. As a result, for all $u$
for which
\begin{equation}
\int\limits^u_{u_0}du\,\kappa <\frac{1}{|\O|}
\end{equation}
one obtains
\begin{equation}
\EH (u)=M(1+\O)\;,\qquad \BH (u)=1+\O\;.
\end{equation}
This is a specific case of eq. (4.46). If eq. (4.46) were valid
for any point $(u,v)$ inside the AH, the result (4.50) would
hold for all $u$.

The remedy is in the fact that $\Gamma_2$ decreases also along
the lines $\v$ including the line $v=v_0$, and, as distinct
from the previous case, this decrease is unbounded. Indeed,
since $u$ is $u^+$ or differs from $u^+$ by a finite
transformation, it may take arbitrarily large values.
As it becomes $u=\uH (v)+M/|\O|$, one obtains owing to conditions
(4.17)
\begin{equation}
\makebox[14cm][l]{$\displaystyle u=\uH (v)+\frac{M}{|\O|}\,\colon
\qquad\qquad
-\Gamma_2=\frac{1}{|\O|}\;,\qquad \ru =\frac{1}{|\O|}\;.$}
\end{equation}
In particular,
\begin{equation}
\makebox[14cm][l]{$\displaystyle u=u_0+\frac{M}{|\O|}\,\colon
\qquad\qquad\qquad\qquad\qquad
\int\limits^u_{u_0}du\,\kappa =\frac{1}{|\O|}\;.$}
\end{equation}
This allows $\EH (u)$ and $\BH (u)$ to differ macroscopically
from their values in eq. (4.50). If the points 4 and 5 are at
that large $u$, then point 6 is already outside the region (4.1).

The above raises the issue of the range of $u$ for which the
presently considered solution is valid. In the region of strong
field, the reserve accompanying eq. (3.12) transfers to the
conditions (4.14) and (4.15) for the data at the AH. It may
turn out that these conditions hold only up to a certain point
of the AH with a finite value of $u^+$. Then the present solution 
is valid only in the causal past of this point. Where the matter
stands is, however, unknown and will be known only at the final
stage when the data functions will be obtained. Therefore, it
makes sense to drive the solution to the limit $u^+\to\infty$
within the present assumptions. It will then be easy to cut
it off at any value of $u^+$.

Eq. (4.51) explains the nature of the bound (4.33). At $v$
fixed, and $u$ as large as in eq. (4.51), this bound is
already almost saturated. It is saturated at the limit
$u\to\infty$, $\v$, if the conditions (4.17) for $\kappa$
hold up to $u\to\infty$. One may even admit that $\kappa (u)$
decreases as $u\to\infty$ but the law of decrease is restricted
to the condition
\begin{equation}
\makebox[14cm][l]{$\displaystyle\v ,\;u\to\infty\,\colon
\qquad\qquad
-\Gamma_2\to\infty\;,\qquad \frac{1}{\kappa (u)}
{\rm e}^{\Gamma_2}\to 0\;.$}
\end{equation}
It suffices that
\begin{equation}
\kappa (u)>O\left(\frac{1}{u}\right)\;,\qquad u\to\infty
\end{equation}
and even that
\begin{equation}
\kappa (u)\ge\frac{\mbox{\small const.}}{u}\;,\quad
\mbox{const.}>1\;,\quad u\to\infty\;.
\end{equation}
Then, at the limit $u\to\infty$ along $\v$, one obtains
\begin{equation}
\makebox[14cm][l]{$\displaystyle\v ,\;u\to\infty\,\colon
\qquad\qquad
-\uv\to\infty\;,\qquad\dr\to 0\;.$}
\end{equation}
It follows that the bound (4.33) is the end of the 
chart $u^+$, and the line
\begin{equation}
\makebox[14cm][l]{$\displaystyle\mbox{EH}\,\colon
\qquad\qquad\qquad\qquad
\rr =-2\beta (v)\frac{\BH (v)}{4\EH (v)\kappa (v)}(1+\O)$}
\end{equation}
is the event horizon. It is easy to check that this line
is null ($u=u_{\mbox{\tiny EH}}=\mbox{const.}$), and eq. (4.56)
implies that, as $u\to u_{\mbox{\tiny EH}}$, the red shift
becomes infinite. At the event horizon,
\begin{equation}
r_{\mbox{\tiny EH}}(v)=\rH (v)\left(1-
\frac{\beta (v)}{2\EH (v)\kappa (v)}(1+\O)\right)
\end{equation}
whence
\begin{equation}
u^-_{\mbox{\tiny EH}}-u^-_0=\frac{2\beta_0}{\kappa_0}(1+\O)\;,
\qquad u^+_{\mbox{\tiny EH}}=\infty\;.
\end{equation}
Of the light rays $\u$, only the ones in the interval of $u$
\begin{equation}
u^-_{\mbox{\tiny EH}}>u^->u^-_0
\end{equation}
cross the AH twice and go out to $\I$. The interval (4.60)
is a subinterval of the semiclassical interval (3.4).

Note that
\begin{equation}
\left.\frac{d\ln\beta\hphantom{u}}{du\hphantom{\ln\beta}}
\right|_{u\to u_0}>0\;,\qquad
\left.\frac{d\ln\beta\hphantom{u}}{du\hphantom{\ln\beta}}
\right|_{u\to\infty}<0\;.
\end{equation}
Here the first inequality follows from the fact that, by
eq. (4.13),
\begin{equation}
\left.\frac{d\ln\beta\hphantom{u}}{du\hphantom{\ln\beta}}
\right|_0=+\infty\;.
\end{equation}
The second follows from the assumption that conditions (4.14)
hold up to $u\to\infty$. Then, at $u\to\infty$, $\EH$ remains
finite, and, therefore,
\begin{equation}
\beta\Bigl|_{u\to\infty}=+0\;.
\end{equation}
Eq. (4.61) suggests that there is a point of the AH where
$\beta$ has a maximum. Call it point~I.
\begin{equation}
\makebox[14cm][l]{$\displaystyle\mbox{point I}\,\colon
\qquad\qquad\qquad\qquad
\frac{d\beta}{du}=0\;,\qquad 
\beta=\beta_{\mbox{\scriptsize max}}\;.$}
\end{equation}
By eq. (4.13),
\begin{equation}
\makebox[14cm][l]{$\displaystyle v<v_{\mbox{\tiny I}}\,\colon
\qquad\qquad\qquad\qquad\qquad
\alpha\frac{d\uH}{dv_{\hphantom{\mbox{\tiny AH}}}}
<\frac{\BH}{4\EH\kappa}\;,$}
\end{equation}
\begin{equation}
\makebox[14cm][l]{$\displaystyle v=v_{\mbox{\tiny I}}\,\colon
\qquad\qquad\qquad\qquad\qquad
\alpha\frac{d\uH}{dv_{\hphantom{\mbox{\tiny AH}}}}
=\frac{\BH}{4\EH\kappa}\;,$}
\end{equation}
\begin{equation}
\makebox[14cm][l]{$\displaystyle v>v_{\mbox{\tiny I}}\,\colon
\qquad\qquad\qquad\qquad\qquad
\alpha\frac{d\uH}{dv_{\hphantom{\mbox{\tiny AH}}}}
>\frac{\BH}{4\EH\kappa}\;.$}
\end{equation}
This helps to complete the spacetime diagram of the semiclassical
region.

Along every ray $\u$ that crosses the AH twice, $\rr$ must have
a minimum at some negative value 
of $\rr\colon\,$ $\rr =\rr_{\mbox{\scriptsize min}}$. These minima make a 
line passing in the interior of the AH. From the identity
\begin{equation}
r\partial_v \rr =\rv^{-1}\left(-2A+\frac{1}{2}B\,\rr\right)
\end{equation}
and the solution above one obtains the equation of the line
of minima:
\begin{equation}
\makebox[14cm][l]{$\displaystyle \rr =\rr_{\mbox{\scriptsize min}}
\,\colon\qquad\qquad
\rr =-2\beta (v)\alpha (v)\frac{d\uH (v)}{dv\hphantom{\uH )(}}
(1+\O)\;.$}
\end{equation}
Since $\rr_{\mbox{\scriptsize min}}\le 0$, one finds that all minima
are at $v\ge v_0$.\footnote{They are, therefore, within the validity
of the solution. In view of this fact, eqs. (4.69) or (4.33) prove
the claim made above that, at $\rr <0$, the chart $u^+$ does not
extend beyond $\rr =-|\O|$.} On the other hand, comparing expression 
(4.69) with the bound (4.33) and using eqs. (4.65)-(4.67) one infers
that all minima 
are at $v<v_{\mbox{\tiny I}}$. At $v=v_{\mbox{\tiny I}}$, the line
of minima crosses the event horizon. Then, along the event 
horizon, $\rr$ first decreases and next, upon passing through the 
minimum at $v=v_{\mbox{\tiny I}}$, increases up to the value
\begin{equation}
\rr_{\mbox{\tiny EH}}\to -0\;,\qquad v\to\vH\Bigl|_{u=\infty}
\end{equation}
reached at a finite or infinite value of $v$
\begin{equation}
\vH (\infty)\stackrel{\rm def}{=}\vH\Bigl|_{u=\infty}\;.
\end{equation}
This follows from eqs. (4.57) and (4.63). At $v=\vH (\infty)$, the
event horizon meets with the apparent horizon. Condition (4.15)
implies that $\vH (\infty)$ is infinite. Then the apparent
horizon is asymptotically tangent to the event horizon. However,
condition (4.15) may not hold up to $u=\infty$, and 
then $\vH (\infty)$ may be finite. In this case, the apparent
horizon either crosses the event horizon or originates its new, 
third, branch. In general, the AH may in the chart $u^+$ have
any even number of branches.

The fact that the line of minima crosses the event horizon means
that the event horizon is not the border of the semiclassical
region. Having obtained the metric in the chart $u^+$, one
should be able to analytically continue it along the rays $\v$
until these rays reach the ultraviolet region. The case of
infinite $\vH (\infty)$ is easy to complete. Consider the function
\begin{equation}
\gamma =-2\alpha\frac{d\EH}{dv_{\hphantom{\mbox{\tiny AH}}}}
=\beta\alpha\frac{d\uH}{dv_{\hphantom{\mbox{\tiny AH}}}}\;.
\end{equation}
If $\vH (\infty)$ is infinite, one has
\begin{equation}
\gamma\Bigl|_{u\to\infty}=+0
\end{equation}
for the same reason for which eq. (4.63) holds. Then there should be
a point of the AH where $\gamma$ has a maximum. Call it point II.
\begin{equation}
\makebox[14cm][l]{$\displaystyle\mbox{point II}\,\colon
\qquad\qquad\qquad\qquad
\frac{d\gamma}{du}=0\;,\qquad 
\gamma =\gamma_{\mbox{\scriptsize max}}\;.$}
\end{equation}
Point II is in the future of point I. This is seen from eqs. 
(4.65)-(4.67): at point I, $\gamma$ is still growing. It follows
from eq. (4.69) that, at $v=v_{\mbox{\tiny II}}$, the line of
minima is tangent to $\u$, and, at $v=\infty$, it meets with
the apparent horizon. At $v>v_{\mbox{\tiny II}}$, this line
is already the line of maxima of $\rr$. The rays $\u$ with
$u>u_{\mbox{\tiny EH}}$ cross it twice (or never cross it).
Along these rays, $\rr$ remains negative and, after the second
crossing, decreases already incessantly pulling these rays
to the ultraviolet region $\rr =-1/|\O|$.

At $u<u_0$, the line of minima recedes into the support of $\t$
and, there, extends to $u=-\infty$. All rays $\u$ that go out
to $\I$ cross this line. The completed spacetime diagram showing
the event horizon and the line of extrema of $\rr$ is given in
Fig. 2 for the case of infinite $\vH (\infty)$.

One more question of interest can be answered with the solution
above. The light rays $\u$ cannot cross the AH the second time
if the energy current through $\u$, $\partial_v E$, is nonnegative.
At least around the second branch of the AH, there should be a band
in which $\partial_v E <0$ and, thereby, the dominant energy
condition is violated. Inserting the solution above in eq. (1.13), 
one obtains at the AH, at the line of minima, and at the line 
$v=v_0$
\begin{eqnarray}
&&(\partial_v E)\Bigl|_{\mbox{\tiny AH}}=-\frac{1}{2}\beta (v)
\BH (v)\frac{d\uH (v)}{dv\hphantom{\uH )(}}\;,\\
&&(\partial_v E)
\Bigl|_{(\nabla r)^2=(\nabla r)^2_{\mbox{\tiny min}}}=
-\frac{1}{2}\beta (v)\frac{d\uH (v)}{dv\hphantom{\uH )(}}
(1+\O)\;,\\
&&(\partial_v E)\Bigl|_{v=v_0}=\beta_0\,\O\;.
\end{eqnarray}
The latter equation follows from the fact that, at $v=v_0$, $B=1+\O$. 
The boundary of the band of negative $\partial_v E$ is
\begin{equation}
\makebox[14cm][l]{$\displaystyle\partial_v E=0\,\colon
\qquad\qquad\qquad
\rr =2\beta (v)\alpha (v)
\frac{d\uH (v)}{dv\hphantom{\uH )(}}\frac{\BH (v)}{1-\BH (v)+\O}\;.$}
\end{equation}

One finds that, at the line of minima and at the second branch
of the AH, $\partial_v E$ is negative, and at the first branch 
positive as it should. Inside the AH, the boundary passes
in CL.I, i.e., in the $O(M)$ neighbourhood of the line $v=v_0$
where the deviation of $\partial_v E$ from zero is within the 
quantum uncertainty. Therefore, at $v=v_0+M/|\O|$, the boundary
$\partial_v E=0$ cannot already get inside the AH. This has
an important consequence for the data at the second branch
of the AH:
\begin{equation}
\BH\le 1+\O\;.
\end{equation}
Indeed, for $v<v_{\mbox{\scriptsize crit}}$, the equality
$\BH =1+\O$ holds by the correspondence principle, and,
for $v>v_{\mbox{\scriptsize crit}}$, condition (4.79) follows
from the fact that $\rr$ in eq. (4.78) cannot be negative.

Eq. (4.78) is valid only when the boundary that it defines is
in the region of strong field. Outside the AH it can be used
if, at sufficiently large $v$, $\BH$ becomes macroscopically
distinct from 1. The band of negative $\partial_v E$ is then
such as shown in Fig. 2. At $\I$, the positivity of
$\partial_v E$ restores but this does not mean that the energy 
dominance restores in full. In fact, the violation of the
dominant energy condition in the region of strong field occurs
with certainty for only one null projection of $\T\,\colon$ $A$. Indeed,
from eqs. (4.29), (4.28), and (4.79) one finds for $v>v_0$
\begin{equation}
A<0\;,\qquad T_1\le\O\;.
\end{equation}
Nonpositivity of $T_1$ is what the dominant energy condition 
requires, and the behaviour of $\partial_v E$ is a consequence
of eq. (4.80). If, outside the AH, $\ru$ decreases along $\u$
monotonically, then the negativity of $A$ persists up to $\I$.
This follows from the identity
\begin{equation}
r\partial_v \ru=-A\rv^{-1}\left(\dr\right)^{-1}
\end{equation}
analogous to (4.8).

Of the four data functions (4.11), two are responsible for the 
choice of $u$ and $v$. Since eqs. (4.7)-(4.10) are invariant
with respect to this choice, the arbitrariness of this choice
in the solution above is restricted only to the conditions
\begin{equation}
\dr =\O\;,\qquad |\O|<\rv<\frac{1}{|\O|}\;.
\end{equation}
Fixing $u$ as $u^+$, and $v$ as in eq. (1.4) will determine two
of the data functions but, for that, the data at the AH should
be related to the data at $\I$ and $\mI$. This requires knowing
the solution globally and not only in the region of strong
field. The remaining two data functions are the basic curvature
scalars $E$ and $B$ at the AH. They too should be related to
the data at the asymptotically flat infinity. Thus one still
needs field equations but, remarkably, only for the region of
weak field.
}

\newpage

{\renewcommand{\theequation}{5.\arabic{equation}}

\begin{center}
\section{\bf    Field equations}
\end{center}

$$ $$

Although, in the semiclassical region, $\T$ need not be small,
some of its projections may remain small. One such projection
one does have. Since it suffices to consider the massless
vacuum particles, the full trace of $\T$ is zero or, at most,
has an anomaly. It does not matter whether there is no anomaly
or there is one, and what is its specific form. It is only 
important that the trace anomaly is local and proportional
to the quantum constant. This is sufficient for the equation
\begin{equation}
T_1+T_2=\O
\end{equation}
to hold throughout the semiclassical region. It can then be
used as one of the field equations.

However, under spherical symmetry, one needs two field equations.
This raises the question if any other projection of $\T$ can
be assumed small throughout the semiclassical region. The answer
is that any such assumption is correct provided that it brings
to a solution, and the solution confirms the assumption. This
follows from the fact that the initial values of all the
projections are small. The problem is to find an assumption of
this kind that would stand its own dynamics. Fortunately,
there is a hint.

The hint is that no assumption may be made about the data at
the AH because these data will subsequently be determined
by the data at infinity, and the data at infinity are subject 
to the quantum dynamics. At the kinematical level, the data
functions should remain arbitrary. Keeping the data at the AH arbitrary,
one can trace $\T$ on going out of the region of strong field
to see if any its projection becomes small. Only such a projection 
can remain small in the region of weak field. A convenient way
of going out of the region of strong field is moving away from
the AH towards earlier $u$ along the lines $\v$

Right away one arrives at an important inference that the 
projection $T_1$ cannot be assumed small. Indeed, if one admits
that the data function $\BH$ may become macroscopically distinct
from 1, then $T_1$ at the AH is not small. Moreover, $T_1$ is
conserved along the lines $\v$ throughout the region of strong 
field and, therefore, does not become small on going out of this
region.

Another example that is worth mentioning is the energy 
current $\partial_v E$. As distinct from $T_1$, $\partial_v E$ at
the AH is small but, on going out of the region of strong
field, it grows and tends to a macroscopic value along 
with $T_1$. Therefore, in the region of weak field, $\partial_v E$
cannot be assumed small either. The projection $T_2$ is expressed
through $T_1$ by eq. (5.1) and is not small if $T_1$ is not small.
About the projection $D$, there is no information except its
boundedness. There remains to be considered the projection $A$,
and here one has good luck. This projection is small at the
AH and is conserved along the lines $\v$ throughout the region
of strong field. It is, therefore, small on going out of this
region and can be assumed small outside this region.

To summarize, there is only one viable candidate for the second
field equation:
\begin{equation}
A=\O\;.
\end{equation}
Of course, at some initial stage of the evolution, all projections
of $\T$ remain small. This stage may be called the epoch of small
vacuum currents but, if $\BH$ breaks away from 1, this epoch has
an end.\footnote{The equation $T_1=\O$ (combined with eq. (5.1)) is
necessary and sufficient for all projections of $\T$ to be $\O$.
The sufficiency follows from the second form of eq. (1.15). Indeed,
if $T_1$ is $\O$, then its first derivative $\partial_v$ is also
$\O$ because its second derivative is bounded (the proof repeats
the one in the Appendix). Besides, the behaviours of $T_1$ and 
$T_2$ at infinity should be taken into account (see the next
section). Then the identity (1.15) takes the form of the following
equation for $A$:
$$
\frac{d\hphantom{r}}{dr}\left(\left.\frac{A}{\rv^2}
\right|_{v=\mbox{\scriptsize const.}}\right)=\frac{\O}{r^2}\;.
$$
With the initial condition $A\bigl|_{\mI}=0$, its solution is
$A=\O$. Finally, as shown below, the equation $A=\O$ combined
with eq. (5.1) implies $D=\O$.} The field equation (5.2)
deliberately covers this epoch and has a chance to outlast it.
The bounds to the validity of this equation are set by conditions
(4.14) and (4.15) for the data functions. As long as these
conditions are valid, $A\bigl|_{\mbox{\tiny AH}}$ in eq. (4.35)
is small. As soon as it will cease being small, the end will come
to the validity of eq. (5.2) as well.

Eq. (5.2) implies that also the first derivatives of $A$ are $\O$.
Specifically, eq. (3.7) gets replaced with
\begin{equation}
M(\partial_u r)^{-1}\partial_u A=\O\;.
\end{equation}
This follows from the boundedness of the second derivatives of
$A$ (see the Appendix).
}

\newpage

{\renewcommand{\theequation}{6.\arabic{equation}}

\begin{center}
\section{\bf    Solution in the region of weak field}
\end{center}

$$ $$

It is not that the field equations (5.1) and (5.2) are valid only
in the region of weak field. They are valid globally but the
accuracy with which they are given enables one to use them only
in the region of weak field. The condition defining this region is
\begin{equation}
-\frac{\partial r^{\hphantom{+}}}{\partial u^+}>|\O|\;.
\end{equation}
In the chart $u^+$, it is equivalent to the condition
\begin{equation}
\rr >|\O|\;.
\end{equation}

In the region of weak field, the initial data to the field equations
cannot be taken at the AH but can be taken at the asymptotically
flat infinity. For the metric at the future null infinity, one
writes the general analytic expansion
\begin{equation}
\rr\|=1-\frac{2{\cal M}(u)}{r}+\frac{Q^2(u)}{r^2}+\cdots\;,
\end{equation}
\begin{equation}
(\nabla r,\nabla u^+)\|=-1-\frac{c_1(u)}{r}-
\frac{c_2(u)}{r^2}+\cdots\;.
\end{equation}
(The coefficient $Q^2(u)$ is not necessarily positive.) Similar
expansions hold for the metric at the past null infinity and
spatial infinity (${\rm i}^0$). The coefficients of these
expansions will be called charges because they represent the
strengths of long-range fields having their sources in a compact
domain. The coefficients of the expansions at $\I$ and $\mI$ (the
Bondi charges) describe respectively the emission and absorbtion
of charges by an isolated system. Thus, the Bondi 
mass ${\cal M}(u)$ is the amount of the gravitational charge that,
in the process of emission, remains in the compact domain by
the instant $u$ of retarded time. Other coefficients may involve
matter charges. For example, if the total $T^{\mu\nu}$ is the
energy-momentum tensor of a system of electric charges and their
electromagnetic field, then $Q^2(u)$ is positive, and $Q(u)$ is
the Bondi electric charge of this system.

The full, or ADM charges appearing as coefficients of the expansion
at ${\rm i}^0$ are the limits as $u^+\to-\infty$ of the Bondi
charges at $\I$. Because $\t$ is assumed to have a compact
spatial support, and $\T$ is retarded, there is no flux of charges
through $\mI$. Therefore, the Bondi charges at $\mI$ are constant and
equal to the respective ADM charges. The ADM charges are conserved,
and, initially, the only nonvanishing macroscopic charge is the
ADM mass $M$ but it would be absurd to exclude a possibility for
the source to have microscopic amounts of other charges, e.g.,
a microscopic electric charge. Therefore,
\begin{equation}
{\cal M}(-\infty)=M\;,\qquad Q^2(-\infty)=\O\;,\qquad\ldots
\end{equation}
\begin{equation}
c_1(-\infty)=\O\;,\qquad c_2(-\infty)=\O\;,\qquad\ldots\;.
\end{equation}

A limitation on the behaviour of the metric at $\I$ stems from the
locality of the trace of $\T$. From eqs. (6.3), (6.4) one can
calculate
\begin{equation}
T_1\|=\frac{c_1}{r}+\cdots\;,\qquad 
T_2\|=-\frac{dc_1^{\hphantom{+}}}{du^+_{\hphantom{1}}}+\cdots\;.
\end{equation}
On the other hand, with these behaviours, all possible local
invariants in the trace anomaly decrease in such a way that
$T_1+T_2$ vanishes at $\I$. Hence
\begin{equation}
\frac{dc_1}{du_{\hphantom{1}}}=0\;,
\end{equation}
and thus, with regard for the initial condition in eq. (6.6),
\begin{equation}
c_1(u)=\mbox{const.}=\O\;.
\end{equation}
With constant $c_1$ one calculates the traces anew and obtains
\begin{equation}
T_1\|=\frac{1}{r}c_1+\frac{1}{r^2}\left(2c_2-c_1^2-
2{\cal M}c_1-Q^2\right)+\cdots\;,
\end{equation}
\begin{equation}
T_2\|=-\frac{1}{r}\left(c_1+
2\,\frac{dc_2^{\hphantom{+}}}{du^+_{\hphantom{2}}}\right)+\cdots
\end{equation}
but the trace anomaly decreases faster:
\begin{equation}
\left(T_1+T_2\right)\|=\frac{\O}{r^3}+\cdots\;.
\end{equation}
Hence
\begin{equation}
\frac{dc_2}{du_{\hphantom{2}}}=0\;,
\end{equation}
\begin{equation}
c_2(u)=\mbox{const.}=\O\;,
\end{equation}
and therefore
\begin{equation}
T_1\|=\frac{c_1}{r}-\frac{Q^2(u)+\O}{r^2}+\cdots\;.
\end{equation}
Eqs. (6.12) and (6.15) are the final results for $T_1$ and
$T_2$ at $\I$.

To summarize, the behaviours of the basic curvature scalars
at $\I$ are
\begin{equation}
E\|={\cal M}(u)-\frac{1}{2}\frac{Q^2(u)}{r}+\cdots\;,
\end{equation}
\begin{equation}
B\|=\frac{2{\cal M}(u)+c_1}{r}-
\frac{2Q^2(u)+\O}{r^2}+\cdots\;.
\end{equation}
The behaviours at $\mI$ and ${\rm i}^0$ are similar
\begin{equation}
E\biggl|_{{\cal I}^-,\ {\rm i}^0}=M+\frac{\O}{r}+\cdots\;,
\end{equation}
\begin{equation}
B\biggl|_{{\cal I}^-,\ {\rm i}^0}=\frac{2M+c_1}{r}+
\frac{\O}{r^2}+\cdots
\end{equation}
but with all the coefficients constant.\footnote{Survival of the 
constant $c_1$ has an alerting consequence at ${\rm i}^0$. Namely,
the radial acceleration of a freely falling particle depends
on its velocity even at the Newtonian limit:
$$
\left.\frac{d^2r}{ds^2}\right|_{{\rm i}^0}=
-\frac{1}{r^2}\left[(M+c_1)+c_1\left(\frac{dr}{ds}\right)^2\right]
+\cdots\;.
$$
Here $s$ is the proper time of the particle. However, since
$c_1=\O$, only a particle of enormous energy can feel this
dependence. Therefore, {\it apriori} this consequence cannot be used
to rule $c_1$ out.}

Eqs. (4.81) and (6.9) have the following consequence for the
curvature $A$:
\begin{equation}
rA\|=\O\;.
\end{equation}
Using the asymptotic behaviours (6.18)-(6.19) in eq. (1.13),
one infers that $A$ has the same property at $\mI$ and ${\rm i}^0$:
\begin{equation}
rA\biggl|_{{\cal I}^-,\ {\rm i}^0}=\O\;.
\end{equation}
This property thus holds as $r\to\infty$ in any direction.
Therefore, the field equation (5.2) can be strengthened as
\begin{equation}
rA=\O\;.
\end{equation}
In the region of weak field, it can now be integrated. For that,
rewrite the identity (4.81) as
\begin{equation}
(\partial_v r)^{-1}\partial_v\ln|\ru|=
-\frac{4A}{r\left(\rr\right)^2}
\end{equation}
and integrate it along $\u$ with the initial condition at $\I$:
\begin{equation}
\ln|(\nabla r,\nabla u^+)|=\int\limits_r^{\infty}
\left.\frac{dr}{r}\frac{4A}{\left(\rr\right)^2}
\right|_{u=\mbox{\scriptsize const.}}\;.
\end{equation}
Insertion of eq. (6.22) and use of condition (6.2) give
\begin{equation}
\makebox[14cm][l]{$\displaystyle\rr>|\O|\,\colon
\qquad\qquad\qquad
\ln|(\nabla r,\nabla u^+)|=\int\limits_r^{\infty}
\frac{dr}{r}\frac{\O}{r}=\O\;.$}
\end{equation}
Hence one obtains the solution in the region of weak field:
\begin{equation}
\makebox[14cm][l]{$\displaystyle\rr>|\O|\,\colon
\qquad\qquad\qquad\qquad
(\nabla r,\nabla u^+)=-1+\O\;.$}
\end{equation}
As a consequence, one has
\begin{equation}
\makebox[14cm][l]{$\displaystyle\rr>|\O|\,\colon
\qquad\qquad\qquad\qquad
\frac{\partial r^{\hphantom{+}}}{\partial u^+}=
-\frac{1}{2}\rr\,(1+\O)\;.$}
\end{equation}

As the next step, rewrite the identity (1.15) (the second form) as
\begin{equation}
r(\partial_v r)^{-1}\partial_v T_1+2T_1=\frac{2}{\rr}
\left[(T_1+T_2)\rr -4AD-2r(\partial_u r)^{-1}\partial_u A\right]\;.
\end{equation}
Here equation (5.3) is to be used. It follows from the asymptotic 
behaviours above that this equation can be strengthened as
\begin{equation}
r(\partial_u r)^{-1}\partial_u A=\O\;.
\end{equation}
Use eqs. (5.1), (5.2), (6.29), and the boundedness of $D$. The
result is that, in the region of weak field, eq. (6.28) closes
as a differential equation along $\u$:
\begin{equation}
\makebox[14cm][l]{$\displaystyle\rr>|\O|\,\colon
\qquad\qquad\qquad\qquad
r(\partial_v r)^{-1}\partial_v T_1+2T_1=\O\;.$}
\end{equation}
Introduce the notation
\begin{equation}
Y=r(\partial_v r)^{-1}\partial_v\left(T_1-\frac{c_1}{r}\right)
+2\left(T_1-\frac{c_1}{r}\right)\;.
\end{equation}
Inserting in eq. (6.31) the analytic expansion of $T_1$ at $\I$,
one finds that the expansion of $Y$ at $\I$ begins with $1/r^3$
because the term of order $1/r^2$ drops out. Then
\begin{equation}
T_1-\frac{c_1}{r}=\frac{1}{r^2}\left(a(u)-
\int\limits_r^{\infty}dr\,
rY\biggl|_{u=\mbox{\scriptsize const.}}\right)
\end{equation}
where $a(u)$ is an arbitrary function of $u$, and the integral
with $Y$ converges. By eqs. (6.30) and (6.31),
\begin{equation}
\makebox[14cm][l]{$\displaystyle\rr>|\O|\,\colon
\qquad\qquad\qquad\qquad
Y=\O-\frac{c_1}{r}=\O$}
\end{equation}
because $c_1=\O$. Hence, using eq. (6.15) to identify $a(u)$
with the data at $\I$, one obtains the solution:
\begin{equation}
\makebox[14cm][l]{$\displaystyle\rr>|\O|\,\colon
\qquad\qquad\qquad\qquad
T_1=\frac{c_1}{r}-\frac{Q^2(u)+\O}{r^2}\;.$}
\end{equation}

As the last step, rewrite the identity (4.68) as
\begin{equation}
r(\partial_v r)^{-1}\partial_v\left(1-\rr\right)+
\left(1-\rr\right)=Z
\end{equation}
with
\begin{equation}
Z=\frac{4A}{\rr}-T_1\;.
\end{equation}
Inserting in eq. (6.35) the analytic expansion of $\rr$ at $\I$,
one finds that the expansion of $Z$ at $\I$ begins with $1/r^2$.
Then
\begin{equation}
1-\rr=\frac{1}{r}\left(b(u)-
\int\limits_r^{\infty}dr\,
Z\biggl|_{u=\mbox{\scriptsize const.}}\right)
\end{equation}
where $b(u)$ is an arbitrary function of $u$, and the integral
with $Z$ converges. By eqs. (6.36), (6.34), (6.9), and (5.2),
in the region of weak field
\begin{equation}
\makebox[14cm][l]{$\displaystyle\rr>|\O|\,\colon
\qquad\qquad\qquad\qquad
Z=\frac{Q^2(u)}{r^2}+\O\;.$}
\end{equation}
Hence, using eq. (6.3) to identify $b(u)$ with the data at $\I$,
one obtains
\begin{equation}
\makebox[14cm][l]{$\displaystyle\rr>|\O|\,\colon
\qquad\qquad\qquad
\rr =1-\frac{2{\cal M}(u)}{r}+\frac{Q^2(u)+\O}{r^2}\;.$}
\end{equation}
Eqs. (6.39) and (6.26) give the solution for the metric in the
region of weak field. For the basic curvature scalars one obtains
\begin{equation}
\makebox[14cm][l]{$\displaystyle\rr>|\O|\,\colon
\qquad\qquad\qquad
E={\cal M}(u)-\frac{1}{2}\frac{Q^2(u)+\O}{r}\;,$}
\end{equation}
\begin{equation}
\makebox[14cm][l]{$\displaystyle\rr>|\O|\,\colon
\qquad\qquad\qquad
B=\frac{2{\cal M}(u)+c_1}{r}-\frac{2Q^2(u)+\O}{r^2}\;.$}
\end{equation}

There remains to be calculated the curvature $D$. For that,
rewrite the identity (1.14) as
\begin{equation}
D\,\rr\dr=2\partial_u E+T_1\dr\;.
\end{equation}
The integral form of this identity is
\begin{equation}
2(M-E)=\int\limits_r^{\infty}dr\left.\left(D\rr-
\frac{c_1}{r}\right)\right|_{v=\mbox{\scriptsize const.}}-
\int\limits_r^{\infty}dr\left.\left(T_1-
\frac{c_1}{r}\right)\right|_{v=\mbox{\scriptsize const.}}\;.
\end{equation}
Insertion of the solution above in eq. (6.42) yields the result
for $D$ in the region of weak field:
\begin{equation}
\makebox[14cm][l]{$\displaystyle\rr>|\O|\,\colon
\qquad\quad
D\,\rr\dr=2\left(\frac{d{\cal M}(u)}{du\hphantom{{\cal M})(}}
-\frac{1}{2r}\frac{dQ^2(u)}{du\hphantom{Q^2)(}}\right)
+\frac{\O}{r}\dr\;.$}
\end{equation}
It is shown in the next section that, with the choice $u=u^+$, the 
derivatives of both data functions ${\cal M}(u)$ and $Q^2(u)$ are
\begin{equation}
\frac{d{\cal M}(u)^{\hphantom{+}}}{du^+\hphantom{{\cal M})(}}=\O\;,\qquad
\frac{dQ^2(u)^{\hphantom{+}}}{du^+\hphantom{Q^2)(}}=\O\;.
\end{equation}
Then, in view of eq. (6.27),
\begin{equation}
\makebox[14cm][l]{$\displaystyle\rr>|\O|\,\colon
\qquad\qquad\qquad\qquad\qquad
D=\O\;.$}
\end{equation}

Eq. (6.46) is analogous to $A=\O$ but the equation analogous to (6.22)
is not true:
\begin{equation}
rD\ne\O\;.
\end{equation}
Indeed, from eq. (6.42) and the asymptotic conditions for $T_1$ one
finds that
\begin{equation}
D\biggl|_{{\cal I}^-,\ {\rm i}^0}=\frac{c_1}{r}+\cdots\;,
\qquad c_1=\O
\end{equation}
but
\begin{equation}
D\|=-4\,\frac{d{\cal M}(u)^{\hphantom{+}}}{du^+\hphantom{{\cal M})(}}
+\cdots\;,
\end{equation}
i.e., $D$ vanishes not in all directions as $r\to\infty$. Owing
to this fact, $E$ can differ macroscopically from its value at
$\mI$ even in the epoch of small vacuum currents, and one can
answer the question where on the incoming light ray $\v$ does
the difference (6.43) accumulate. It accumulates at large $r$,
$r=M/|\O|$, but not in the asymptotic region of $\mI$. It 
accumulates only on very late rays, $v-v_0=M/|\O|$, because,
after having passed the region of $\mI$, these rays get into
the region of validity of eq. (6.49), i.e., into the
asymptotic region of $\I$. In this region, their $r$ is of order
\begin{equation}
r=MO\left(\frac{d{\cal M}(u)^{\hphantom{+}}}{du^+
\hphantom{{\cal M})(}}\right)^{-1}\;,
\end{equation}
and the deficit of $E$ that accumulates when passing across
this region is
\begin{equation}
M-E=M-{\cal M}(u)+\O\;.
\end{equation}
When the ray $\v$ goes out into the region $r=O(M)$, the variation
of $E$ along it can continue only if $Q^2(u)$ is already macroscopic.
Otherwise, this variation ceases, and it will be recalled that
no such variation occurs when approaching the AH. On the line
$\v$, the AH is a regular point. The variation of $E$ along
$\v$ is an effect of weak field but this effect is responsible for 
the distinction of $\EH (v)$ from $M$.

Eq. (6.47) is also the cause for which the calculation of $\rv$
is more involved than the calculation of $\ru$ in eqs. 
(6.23)-(6.26). Integrating eq. (4.8) with the initial condition
at $\mI$, one obtains
\begin{equation}
\ln\rv=\int\limits_r^{\infty}\left.\frac{dr}{r}D
\right|_{v=\mbox{\scriptsize const.}}\;,
\end{equation}
and insertion of expression (6.44) yields
\begin{equation}
\makebox[14cm][l]{$\displaystyle\rr>|\O|\,\colon
\qquad
\ln\rv=2\int\limits_r^{\infty}\left.\frac{dr}{r}
\frac{d{\cal M}(u)^{\hphantom{+}}}{du^+\hphantom{{\cal M})(}}
\left(\rr\frac{\partial r^{\hphantom{+}}}{\partial u^+}
\right)^{-1}\right|_{v=\mbox{\scriptsize const.}}+
\frac{\O}{r}\;.$}
\end{equation}
Since $d{\cal M}/du^+=\O$, the sector of the integration path at
which $r=O(M)$ contributes~$\O$:
\begin{equation}
\makebox[14cm][l]{$\displaystyle\rr>|\O|\,\colon
\qquad
\ln\rv=\O+2\int\limits^{\infty}_{M/|\O|}\left.\frac{dr}{r}
\frac{d{\cal M}(u)^{\hphantom{+}}}{du^+\hphantom{{\cal M})(}}
\left(\rr\frac{\partial r^{\hphantom{+}}}{\partial u^+}
\right)^{-1}\right|_{v=\mbox{\scriptsize const.}}\!\!\!.$}
\end{equation}
In the remaining integral, one can put $\rr=1+\O$ to obtain
\begin{eqnarray}
\int\limits^{\infty}_{M/|\O|}\left.\frac{dr}{r}
\frac{d{\cal M}(u)^{\hphantom{+}}}{du^+\hphantom{{\cal M})(}}
\left(\frac{\partial r^{\hphantom{+}}}{\partial u^+}\right)^{-1}
\right|_{v=\mbox{\scriptsize const.}}&\!=\!&
\frac{|\O|}{M}\int\limits^{\infty}_{M/|\O|}dr\,\left.
\frac{d{\cal M}(u)^{\hphantom{+}}}{du^+\hphantom{{\cal M})(}}
\left(\frac{\partial r^{\hphantom{+}}}{\partial u^+}\right)^{-1}
\right|_{v=\mbox{\scriptsize const.}}\nonumber\\&\!=\!&
\frac{|\O|}{M}(M-{\cal M})=\O\;.
\end{eqnarray}
This calculation can be done more rigorously by dividing the
integration interval in eq. (6.52) into three: $\rr=\O$,
$|\O|\!<\!\rr\!<\! 1-|\O|$, and $\rr=1-|\O|$ with appropriately chosen
border points. The main thing is that the integral (6.52)
contains an extra $1/r$ as compared to the analogous integral
in eq. (6.43). The result is
\begin{equation}
\makebox[14cm][l]{$\displaystyle\rr>|\O|\,\colon
\qquad\qquad\qquad\qquad
\rv=1+\O\;.$}
\end{equation}

The specific form of the $\O$ in expression (6.56) is of interest
only at infinity. To obtain the behaviour of $\rv$ at $\I$,
introduce in eq. (6.53) the integration variable $u$ and go
over to the limit of $\I$ in the integrand using that, by
eq. (4.37),
\begin{equation}
r\|=\frac{v}{2}+\cdots\;.
\end{equation}
The resultant behaviour is
\begin{equation}
\rv\|=1+2\,\frac{M-{\cal M}(u)+\O}{r}+\cdots\;,
\end{equation}
and the behaviours
\begin{equation}
\rv\biggl|_{{\cal I}^-,\ {\rm i}^0}=1+\frac{c_1}{r}+\cdots
\end{equation}
follow from eqs. (6.52) and (6.48).
}

\newpage

{\renewcommand{\theequation}{7.\arabic{equation}}

\begin{center}
\section{\bf    Global solution}
\end{center}

$$ $$

By the consideration above, the quantity (6.22) is uniformly
bounded with some $\O$:
\begin{equation}
\frac{r}{M}|A|<{\bf\cal A}\;,\qquad {\bf\cal A}=\mbox{const.}=\O\;.
\end{equation}
Then, inspecting the right-hand sides of eqs. (6.23), (6.28),
(6.35), and using the result in the Appendix, one infers that
the condition of validity of the weak-field solution is
\begin{equation}
\rr>O(\sqrt{{\bf\cal A}})\;.
\end{equation}
In all the equations (6.25)-(6.44) one can replace the condition
$\rr>|\O|$ with condition (7.2). This implies that there exists
a region:
\begin{equation}
\makebox[14cm][l]{$\displaystyle\mbox{OVERLAP}\,\colon
\qquad\qquad\qquad\qquad
\O=\rr>O(\sqrt{{\bf\cal A}})\;,$}
\end{equation}
or, equivalently,
\begin{equation}
\makebox[14cm][l]{$\displaystyle\mbox{OVERLAP}\,\colon
\qquad\qquad\qquad\qquad
\O=\weak$}
\end{equation}
in which both the weak-field and strong-field solutions are
valid.

The fact that the regions of weak field and strong field overlap
will now be used to relate the data at the AH to the data
at infinity. The data functions $\beta$ and $\kappa$ in
eqs. (4.12) and (4.13) depend on the choice of $u$ for the
strong-field solution. The choice will now be made as $u=u^+$.
The functions $\beta$ and $\kappa$ below refer to this choice.
As far as the data at infinity are concerned, the consistency
requirements in the asymptotic domain bring to no limitations
on the functions ${\cal M}(u)$ and $Q^2(u)$. Only their
initial values are fixed as in eq. (6.5). These values enable
one to use the conditions
\begin{equation}
|\O|<{\cal M}(u)<\frac{1}{|\O|}\;,\qquad 
|Q^2(u)|<\frac{1}{|\O|}
\end{equation}
on the same grounds as conditions (3.12). It will be shown below
that additional limitations on the data functions ${\cal M}(u)$
and $Q^2(u)$ stem from the requirement of boundedness of the
curvature in the compact domain.

First note that eqs. (4.10) and (3.12) enable one to integrate
any equation of the form
\begin{equation}
r\partial_v X=O\left(\dr\right)\;.
\end{equation}
With the data for $X$ on an arbitrary line
\begin{equation}
\makebox[14cm][l]{$\displaystyle L\,\colon
\qquad\qquad\qquad\qquad\qquad\qquad\qquad
v=f(u)\;,$}
\end{equation}
one obtains
\begin{equation}
X=X\left.\vphantom{\dr}\right|_L-O\left(\left.\dr\right|_L\right)+
O\left(\dr\right)\;.
\end{equation}
The solution in this form is valid globally but is useful only
in the region of strong field. Introduce the function
\begin{equation}
K=-(\partial_u r)^{-1}\partial_u\drplus\;.
\end{equation}
Using eqs. (4.10) and (4.8), one can calculate
\begin{equation}
r\partial_v K=-\frac{1}{2\rv}\left[(\partial_u r)^{-1}\partial_u B
+\frac{B(D-1)}{r}\right]\drplus\;,
\end{equation}
and hence, by the boundedness conditions,
\begin{equation}
r\partial_v K=\frac{O(1)}{M}\drplus=\frac{1}{M}\,
O\left(\drplus\right)\;.
\end{equation}
This equation is of the form (7.6). Therefore, the solution
(7.8) applies and, for the region of strong field, yields
\begin{equation}
\makebox[14cm][l]{$\displaystyle\strong\,\colon
\qquad\qquad
K=K\left.\vphantom{\dr}\right|_L-\frac{1}{M}\,O\left(\left.
\drplus\right|_L\right)+\frac{1}{M}\,O\left(\drplus\right)\;.$}
\end{equation}
On the other hand, in the region of weak field one has eq. (6.27).
Combining it with the identity
\begin{equation}
r\partial_u\rr=\left(B-D\rr\right)\dr\;,
\end{equation}
one obtains in the region of weak field
\begin{equation}
\makebox[14cm][l]{$\displaystyle\weak\,\colon
\qquad\qquad\qquad
K=\frac{1}{2r}\left(B-D\rr\right)$}
\end{equation}
and, hence, in the region of overlap
\begin{equation}
\makebox[14cm][l]{$\displaystyle\mbox{OVERLAP}\,\colon
\qquad\qquad\qquad\qquad\qquad
K=\frac{B}{4E}+\O\;.$}
\end{equation}
Eq. (7.15) provides the initial condition for the region of
strong field. Choosing the line $L$ as passing in the region 
of overlap, for example,
\begin{equation}
\makebox[14cm][l]{$\displaystyle L\,\colon
\qquad\qquad\qquad\qquad\qquad\qquad
-\drplus={\bf\cal A}^{\varepsilon}\;,\qquad 
0<\varepsilon<\frac{1}{2}\;,\quad
\varepsilon =\mbox{const.}\;,$}
\end{equation}
\begin{equation}
\left.\frac{dv^{\hphantom{+}}}{du^+}\right|_L=\rv\;,
\end{equation}
one obtains from eqs. (7.12) and (7.15)
\begin{equation}
\makebox[14cm][l]{$\displaystyle\strong\,\colon
\qquad\qquad\qquad\qquad\qquad
K=\left.\frac{B}{4E}\right|_L+\O\;.$}
\end{equation}
Hence
\begin{equation}
\makebox[14cm][l]{$\displaystyle\strong\,\colon
\qquad\qquad\qquad\qquad\qquad
|\O|<K<\frac{1}{|\O|}$}
\end{equation}
in virtue of eq. (3.12).

Next, calculate the action of the operator 
$\partial/\partial u^+$ on the quantity (7.10), and use
condition (7.19) and the boundedness conditions to obtain
\begin{equation}
\makebox[14cm][l]{$\displaystyle\strong\,\colon
\qquad\qquad\qquad
r\partial_v\frac{\partial K}{\partial u^+}
=\frac{O(1)}{M^2}\drplus =\frac{1}{M^2}\,O\left(\drplus\right)\;.$}
\end{equation}
Provided that the line $L$ passes in the region of strong field, 
the solution (7.8) applies again:
\begin{equation}
\makebox[14cm][l]{$\displaystyle\strong\,\colon
\qquad\qquad
\frac{\partial K}{\partial u^+}
=\left.\frac{\partial K}{\partial u^+}\right|_L
-\frac{1}{M^2}\,O\left(\left.\drplus\right|_L\right)
+\frac{1}{M^2}\,O\left(\drplus\right)\;.$}
\end{equation}
On the other hand, in the region of overlap one can differentiate
eq. (7.15):
\begin{equation}
\makebox[14cm][l]{$\displaystyle\mbox{OVERLAP}\,\colon
\qquad\qquad\qquad
\frac{\partial K}{\partial u^+}
=\frac{O(1)}{M^2}\drplus =\frac{1}{M^2}\,O\left(\drplus\right)\;.$}
\end{equation}
Hence, choosing the line $L$ as in eq. (7.16), one obtains
\begin{equation}
\makebox[14cm][l]{$\displaystyle\strong\,\colon
\qquad\qquad\qquad\qquad\qquad
\frac{\partial K}{\partial u^+}=\O\;.$}
\end{equation}
In view of condition (7.19), the same is true of any power
of $K$, e.g.,
\begin{equation}
\makebox[14cm][l]{$\displaystyle\strong\,\colon
\qquad\qquad\qquad\qquad\qquad
\frac{\partial\hphantom{u^+}}{\partial u^+}\frac{1}{K}=\O\;.$}
\end{equation}

Eqs. (7.9) and (7.19) enable one to integrate in the region
of strong field any equation of the form
\begin{equation}
r\partial_u Y=O\left(\dr\right)\;.
\end{equation}
With the data for $Y$ on an arbitrary line
\begin{equation}
\makebox[14cm][l]{$\displaystyle{\cal L}\,\colon
\qquad\qquad\qquad\qquad\qquad\qquad\qquad
u=f(v)$}
\end{equation}
passing in the region of strong field, one obtains
\begin{equation}
\makebox[14cm][l]{$\displaystyle\strong\,\colon
\qquad\qquad\qquad
Y=Y\left.\vphantom{\dr}\right|_{\cal L}
-O\left(\left.\drplus\right|_{\cal L}\right)
+O\left(\drplus\right)\;.$}
\end{equation}
In particular, the boundedness condition (3.11) solves as
\begin{equation}
\makebox[14cm][l]{$\displaystyle\strong\,\colon
\qquad\qquad
\partial_v B=\partial_v B\left.\vphantom{\dr}\right|_{\cal L}
-\frac{1}{M}\,O\left(\left.\drplus\right|_{\cal L}\right)
+\frac{1}{M}\,O\left(\drplus\right)\;.$}
\end{equation}
On the other hand, in the region of weak field one can 
differentiate the solution (6.41):
\begin{equation}
\makebox[14cm][l]{$\displaystyle\weak\,\colon
\qquad\quad
\partial_v B =\frac{\rr}{\rv}\left(-\frac{{\cal M}(u)}{r^2}
+2\,\frac{Q^2(u)}{r^3}+\O\right)\;.$}
\end{equation}
Hence in the region of overlap
\begin{equation}
\makebox[14cm][l]{$\displaystyle\mbox{OVERLAP}\,\colon
\qquad\qquad\qquad\qquad
\partial_v B =\frac{1}{M}\,O\left(\drplus\right)$}
\end{equation}
where use is made of eq. (6.27) and conditions (7.5).
The line ${\cal L}$ can be chosen as $L$ in eq. (7.16).
Indeed, by eq. (7.17), the equation of the line $L$ is
solvable with respect to $u$ as well as with respect to $v$.
As a result, from eqs. (7.28) and (7.30) one obtains
\begin{equation}
\makebox[14cm][l]{$\displaystyle\strong\,\colon
\qquad\qquad\qquad\qquad\qquad
\partial_v B =\O\;.$}
\end{equation}
Then, for $B$ at the AH, one may write
\begin{equation}
\frac{d\BH\hphantom{v}}{dv\hphantom{\BH}}=
(\partial_v B)\Bigl|_{\mbox{\tiny AH}}+
\frac{d\uH\hphantom{v}}{dv\hphantom{\uH}}
(\partial_u B)\Bigl|_{\mbox{\tiny AH}}
\end{equation}
and use eqs. (7.31), (4.15), and (3.9). In this way for the
data at the AH one obtains the condition
\begin{equation}
\frac{d\BH\hphantom{v}}{dv\hphantom{\BH}}=\O
\end{equation}
and, thereby, confirms the assumption (4.16).

The line ${\cal L}$ in eq. (7.26) can, alternatively, be
chosen as the AH. Then, since eqs. (4.7)-(4.9) are of the
form (7.25), they solve as
\begin{equation}
\makebox[14cm][l]{$\displaystyle\strong\,\colon
\qquad\qquad\quad
B=\BH (v)-O\left(\beta(v)\right)+O\left(\drplus\right)\;,$}
\end{equation}
\begin{equation}
\makebox[14cm][l]{$\displaystyle\hphantom{\strong\,\colon
\qquad\qquad\quad}
\rv=\alpha(v)\left[1-O\left(\beta(v)\right)+
O\left(\drplus\right)\right]\;,$}
\end{equation}
\begin{equation}
\makebox[14cm][l]{$\displaystyle\hphantom{\strong\,\colon
\qquad\qquad\quad}
r=2\EH(v)\left[1-O\left(\beta(v)\right)+
O\left(\drplus\right)\right]\;.$}
\end{equation}
With these expressions, eq. (4.10) solves as
\begin{eqnarray}
\makebox[14cm][l]{$\displaystyle\strong\,\colon\qquad
-\left(1+O\left(\drplus\right)\right)\drplus =
\Bigl(1+O\left(\beta(u)\right)\Bigr)\beta(u)$}\nonumber\\
{}\times\exp\left[\int\limits^v_{\vH (u)}dv
\frac{\BH (v)}{4\alpha(v)\EH (v)}
\Bigl(1+O\left(\beta(v)\right)\Bigr)\right]\;.
\end{eqnarray}
Hence
\begin{equation}
\makebox[14cm][l]{$\displaystyle\strong\,\colon
\qquad\qquad\qquad\quad
-\frac{\partial\hphantom{u^+}}{\partial u^+}\drplus
=\Bigl(\kappa(u)+\O\Bigr)\drplus$}
\end{equation}
with the function $\kappa$ in eq. (4.13). From eqs. (7.9) and
(7.38) one infers
\begin{equation}
\makebox[14cm][l]{$\displaystyle\strong\,\colon
\qquad\qquad\qquad\qquad\qquad
K=\kappa(u)+\O\;,$}
\end{equation}
and then eqs. (7.19) and (7.24) yield for the data at the AH
the conditions
\begin{equation}
|\O|<\;\kappa\;<\frac{1}{|\O|}\;,\qquad
\frac{d\hphantom{u^+}}{du^+}\frac{1}{\kappa}=\O\;.
\end{equation}
Thereby, one confirms the assumptions (4.17).

In the region of weak field, differentiate again the solution 
(6.41), this time with respect to $u$:
\begin{eqnarray}
\makebox[14cm][l]{$\displaystyle\weak\,\colon\qquad\qquad
\partial_u B=-\frac{2}{r}\left(\frac{{\cal M}(u)}{r}
-\frac{2Q^2(u)}{r^2}+\O\right)\partial_u r$}\nonumber\\
{}+\frac{2}{r}\left(\frac{d{\cal M}(u)}{du\hphantom{{\cal M})(}}
-\frac{1}{r}\frac{dQ^2(u)}{du\hphantom{Q^2)(}}\right)\;.\qquad\qquad
\end{eqnarray}
By the boundedness conditions (3.9) and (7.5), this relation can be
written in the form
\begin{equation}
\makebox[14cm][l]{$\displaystyle\weak\,\colon\qquad\qquad
\frac{1}{r}\left(\frac{d{\cal M}(u)}{du\hphantom{{\cal M})(}}
-\frac{1}{r}\frac{dQ^2(u)}{du\hphantom{Q^2)(}}\right)
=\frac{O(1)}{M}\dr\;.$}
\end{equation}
From eq. (6.44) and the boundedness of $D$, one obtains another
such relation:
\begin{equation}
\makebox[14cm][l]{$\displaystyle\weak\,\colon\qquad\qquad
\frac{d{\cal M}(u)}{du\hphantom{{\cal M})(}}
-\frac{1}{2r}\frac{dQ^2(u)}{du\hphantom{Q^2)(}}
=O(1)\dr\;.$}
\end{equation}
The point is that the region of validity of these relations 
includes the region of overlap, and, there, 
$\partial r/\partial u^+=\O$. Hence, for all $u^+$ for which
the rays $\u$ reach the region of overlap, one obtains the
conditions
\begin{equation}
\frac{d{\cal M}(u)}{du^+\hphantom{{\cal M}}}
=\O\;,\qquad
\frac{dQ^2(u)}{du^+\hphantom{Q^2}}=\O
\end{equation}
limiting the data at $\I$. Thereby, one proves eq. (6.45).
This limitation, like eqs. (7.33) and (7.40), is a condition
of the existence of the global solution.

Eq. (7.44) is not the only limitation on the data at $\I$
that follows from the consistency requirements in the compact 
domain. The requirement that the weak-field solution be
consistent in the region of overlap and that, moreover,
conditions (3.12) be fulfilled:
\begin{equation}
\makebox[14cm][l]{$\displaystyle\mbox{OVERLAP}\,\colon
\qquad\qquad\qquad\qquad\qquad
rB>M|\O|$}
\end{equation}
brings via eqs. (6.39), (6.41), and (7.5) to the following
limitation on the data functions:
\begin{equation}
\frac{Q^2(u)}{{\cal M}^2(u)}<1-|\O|\;.
\end{equation}
For sufficiently early $u$, conditions (7.44) hold by the
correspondence principle, and condition (7.46) is fulfilled
in consequence of the initial conditions (6.5). If it will
turn out that they are valid only up to some finite value
of $u^+$, then the line $\u$ with this value of $u^+$ bounds
the region of validity of the present solution.

Condition (3.12) for $B$ used in eqs. (6.40) and (6.41) implies
also that in the region of weak field
\begin{equation}
\makebox[14cm][l]{$\displaystyle\weak\,\colon
\qquad\qquad\qquad
E>\frac{1}{2}{\cal M}(u)+|\O|\;.$}
\end{equation}
Then consider the weak-field solution in the region of overlap.
In this region one has $r=2E(1+\O)$, and, therefore, eq. (6.40)
becomes the following equation for $E$:
\begin{equation}
\makebox[14cm][l]{$\displaystyle\mbox{OVERLAP}\,\colon
\qquad\qquad\qquad\qquad
E={\cal M}(u)-\frac{Q^2(u)}{4E}+\O\;.$}
\end{equation}
Condition (7.47) singles out the solution
\begin{equation}
\makebox[14cm][l]{$\displaystyle\mbox{OVERLAP}\,\colon
\qquad\qquad
E=\frac{1}{2}\left({\cal M}(u)+\sqrt{{\cal M}^2(u)-Q^2(u)+\O}
\right)\;.$}
\end{equation}
Hence for $B$ one obtains
\begin{equation}
\makebox[14cm][l]{$\displaystyle\mbox{OVERLAP}\,\colon
\qquad\qquad
B=2-\frac{2{\cal M}(u)}{{\cal M}(u)+\sqrt{{\cal M}^2(u)
-Q^2(u)+\O}}+\O\;.$}
\end{equation}
Finally, in the region of weak field one has eqs. (6.56)
and (6.26) whence
\begin{equation}
\makebox[14cm][l]{$\displaystyle\mbox{OVERLAP}\,\colon
\qquad\qquad
\rv =1+\O\;,\qquad 
(\nabla r,\nabla u^+)=-1+\O\;.$}
\end{equation}

Consider now the strong-field solution. Any path connecting
the AH with the asymptotically flat infinity crosses the region
of overlap. The pathes $\v$ cross it when
\begin{equation}
\makebox[14cm][l]{$\displaystyle\mbox{OVERLAP}\,\colon
\qquad\qquad\qquad
O\left(\frac{1}{\beta(v)}\right)>{\rm e}^{\Gamma_2}
>O\left(\frac{1}{\sqrt{\beta(v)}}\right)\;.$}
\end{equation}
This is seen from eqs. (7.4), (4.29), and (4.20). Therefore,
\begin{equation}
\makebox[14cm][l]{$\displaystyle\mbox{OVERLAP}\,\colon
\qquad\qquad\qquad\qquad\qquad
\Gamma_2=\frac{1}{|\O|}\;.$}
\end{equation}
Using eq. (7.53) and eqs. (4.46)-(4.47), one obtains from
the strong-field solution (4.20)-(4.29)
\begin{equation}
\makebox[14cm][l]{$\displaystyle\mbox{OVERLAP}\,\colon
\qquad\qquad\qquad
\rv=\alpha(v)(1+\O)\;,$}
\end{equation}
\begin{equation}
\makebox[14cm][l]{$\displaystyle\hphantom{\mbox{OVERLAP}\,\colon
\qquad\qquad\qquad}
\ru=-\frac{\BH (u)}{4\EH (u)\kappa(u)}(1+\O)\;,$}
\end{equation}
\begin{equation}
\makebox[14cm][l]{$\displaystyle\hphantom{\mbox{OVERLAP}\,\colon
\qquad\qquad\qquad}
E=\EH (u)(1+\O)\;,$}
\end{equation}
\begin{equation}
\makebox[14cm][l]{$\displaystyle\hphantom{\mbox{OVERLAP}\,\colon
\qquad\qquad\qquad}
B=\BH (u)+\O\;.$}
\end{equation}

Equating the functions in eqs. (7.54)-(7.57) and (7.49)-(7.51)
relates the data at the AH to the data at infinity:
\begin{equation}
\alpha=1+\O\;,
\end{equation}
\begin{equation}
\kappa=\frac{\BH}{4\EH}(1+\O)\;,
\end{equation}
\begin{equation}
\EH (u)=\frac{1}{2}\left({\cal M}(u)+\sqrt{{\cal M}^2(u)
-Q^2(u)+\O}\right)+\O\;,
\end{equation}
\begin{equation}
\BH (u)=2-\frac{2{\cal M}(u)}{{\cal M}(u)+\sqrt{{\cal M}^2(u)
-Q^2(u)+\O}}+\O\;.
\end{equation}
Relations (7.58) and (7.59) result from fixing the
normalizations of $v$ and $u$. In particular, the choice
$u=u^+$ results in the determination of $\kappa$ as in
eq. (7.59). Insertion of expression (4.13) in eq. (7.59)
yields the differential constraint
\begin{equation}
\frac{d\ln\beta}{du^+}=
\frac{\BH}{4\EH}\left(\frac{1}{\alpha}
\frac{d\vH^{\hphantom{+}}}{du^+_{\hphantom{\tiny AH}}}
-1+\O\right)\;,
\qquad\beta=-2\,
\frac{d\EH^{\hphantom{+}}}{du^+_{\hphantom{\tiny AH}}}
\end{equation}
which, combined with eq. (7.58), leaves two independent data
functions at the AH: $\EH$ and $\BH$. These are related to
the data at $\I$ by eqs. (7.60) and (7.61). In consequence of
relation (7.61), condition (4.79) imposes a new limitation on
the data at $\I$:
\begin{equation}
-|\O|\le\frac{Q^2(u)}{{\cal M}^2(u)}\;.
\end{equation}
The inequalities (7.63) and (7.46) clutch the ratio
$Q^2/{\cal M}^2$.

Setting $u=u^+$ and using relations (7.58), (7.59) simplify
the equations of section 4, and it will be added that these
equations enable one to calculate the red-shift factor.
Inserting the solution (4.22) in eq. (3.17), one obtains
\begin{equation}
\frac{du^-}{du^+}=2\beta_0\exp\left(
-\int\limits^{u^+}_{u^+_0}du^+\kappa(u)\right)\;,
\end{equation}
\begin{equation}
\frac{d\hphantom{u^+}}{du^+}\ln\frac{du^+}{du^-}=
\frac{\BH (u)}{4\EH (u)}(1+\O)\;.
\end{equation}

For the curvature $D$ in the region of strong field, integration
of eq. (3.8) yields the result similar to (4.27)
\begin{equation}
\makebox[14cm][l]{$\displaystyle\strong\,\colon
\qquad\qquad\qquad\qquad
D=D_{\mbox{\tiny AH}}(v)+\O$}
\end{equation}
but, within the strong-field solution, one is unable to express
$D_{\mbox{\tiny AH}}$ through the independent data. Adjoining
the weak-field solution (6.46) through the region of overlap,
one obtains $D_{\mbox{\tiny AH}}$ and infers that the equation
\begin{equation}
D=\O
\end{equation}
holds globally. The equations (5.1) and (5.2) also hold globally.
Thus, in the basis (1.8)-(1.11), only one projection of $\T$ can
become macroscopic in the semiclassical region:~$T_1$.
}

\newpage

{\renewcommand{\theequation}{8.\arabic{equation}}

\begin{center}
\section{\bf    Global solution (continued)}
\end{center}

$$ $$

Missing now are analytic expressions for the solution that
could be used in both the weak-field and strong-field regions.
Outside the AH, such expressions can be obtained. Of the main
interest is the global solution for an outgoing light ray. It
can be obtained as follows.

In the region of weak field one can integrate the identity
(4.37) by making use of eq. (6.39). However, for making use
also of eq. (6.56), $r$ should be not large: $r<M/|\O|$ because,
for large $r$, the correction in eq. (6.58) cannot be discarded.
The initial condition can be taken at the line $L$ in the
region of overlap. Then, for
\begin{equation}
\rr\Bigl|_L\;\le\rr<1-|\O|\;,
\end{equation}
one obtains
\begin{eqnarray}
\left(v-v\Bigl|_L\right)=2r+4{\cal M}(u)\ln\frac{r}{2\EH (u)}+
\frac{4\EH (u)}{\BH (u)}\ln
\frac{(\nabla r)^2\hphantom{\Bigl|_L}}{(\nabla r)^2\Bigl|_L}
-4\EH (u)(1+\O)\qquad\nonumber\\
{}-\frac{4\EH (u)}{\BH (u)}\left[1+\Bigl(1-\BH (u)\Bigr)^2\right]
\left\{\ln\left[1-\frac{2\EH (u)}{r}\Bigl(1-\BH (u)\Bigr)\right]-
\ln\BH (u)\right\}
\end{eqnarray}
where the identifications (7.60), (7.61) are used.

In the region of strong field one has eqs. (4.24) and (4.18) in 
which one may insert the expression (7.59) for $\kappa$. Outside
the AH, one may use also eqs. (4.46)-(4.47) to obtain
\begin{equation}
\makebox[14cm][l]{$\displaystyle\rr=|\O|\,\colon
\qquad\qquad\quad
(\nabla r)^2 +2\beta(v)(1+\O)=2\beta(u)\,{\rm e}^{\Gamma_1}(1+\O)\;.$}
\end{equation}
It follows from the derivation of eqs. (4.46)-(4.47) that,
for a point $(u,v)$ outside the AH, the integrand of $\Gamma_1$
in eq. (4.19) is constant up to $\O$:
\begin{equation}
\makebox[14cm][l]{$\displaystyle\rr=|\O|\,\colon
\qquad\qquad\qquad
\Gamma_1=\frac{\BH (u)}{4\EH (u)}\Bigl(v-\vH (u)\Bigr)(1+\O)\;.$}
\end{equation}
Hence one obtains
\begin{equation}
\makebox[14cm][l]{$\displaystyle\rr=|\O|\,\colon
\qquad
(1+\O)\Bigl(v-\vH (u)\Bigr)=\frac{4\EH (u)}{\BH (u)}
\ln\frac{(\nabla r)^2 +2\beta(v)}{2\beta(u)}+\O\;.$}
\end{equation}
When the point $(u,v)$ is at the AH, $\beta(v)=\beta(u)$. In this
way eq. (8.5) verifies at the AH. The solution (8.5) is valid
on the line $L$ as well:
\begin{equation}
(1+\O)\left(v\Bigl|_L-\vH (u)\right)=\frac{4\EH (u)}{\BH (u)}
\ln\frac{(\nabla r)^2\Bigl|_L}{2\beta (u)}+\O\;.
\end{equation}
Here use is made of the fact that
\begin{equation}
(\nabla r)^2\Bigl|_L\;>O\left(\beta^{1/2}(v)\right)\;.
\end{equation}

The bounds (8.1) and (8.7) can be used to replace in equation
(8.2) $(\nabla r)^2$ 
with ${(\nabla r)^2 +2\beta(v)}$ 
within the accuracy of this 
equation. Then, combining eqs. (8.2) and (8.6), one obtains
\begin{eqnarray}
&{\displaystyle\!\!\!\!
(1+\O)\Bigl(v-\vH (u)\Bigr)=2r+4{\cal M}(u)\ln\frac{r}{2\EH (u)}
+\frac{4\EH (u)}{\BH (u)}\ln\frac{(\nabla r)^2 +2\beta(v)}{2\beta(u)}
-4\EH (u)}&\nonumber\\
&{\displaystyle
{}-\frac{4\EH (u)}{\BH (u)}\left[1+\Bigl(1-\BH (u)\Bigr)^2\right]
\left\{\ln\left[1-\frac{2\EH (u)}{r}\Bigl(1-\BH (u)\Bigr)\right]
-\ln\BH (u)\right\}}&
\end{eqnarray}
and concludes that this expression is valid in the whole of
the range
\begin{equation}
0\le\rr<1-|\O|\;.
\end{equation}
Indeed, in the range (8.1) it is valid by derivation, and in
the range $\rr=|\O|$ it coincides with expression (8.5). Outside
the AH, only in the asymptotic region $\rr=1-|\O|$ does the
expression (8.8) need a correction. In this region, one may
use the asymptotic behaviours (6.3) and (6.58) to obtain
the solution of eq. (4.37):
\begin{equation}
\Bigl(v-\vH (u)\Bigr)\|=2r+4M\ln\frac{r}{2\EH (u)}+\cdots\;.
\end{equation}
The correction is thus in the fact that, when $\ln r$ is large,
its coefficient is constant and equal to $4M$ as distinct from 
$4{\cal M}(u)$ in eq. (8.8). With this correction, eq. (8.8)
is the soughtafter solution for an outgoing light ray 
at $\rr\ge 0$.

Outside the AH, also the following relations are valid globally:
\begin{equation}
\makebox[14cm][l]{$\displaystyle\rr\ge 0\,\colon
\qquad\qquad\qquad\quad
-2(1+\O)\drplus =(\nabla r)^2+2\beta(v)\;,$}
\end{equation}
\begin{equation}
\makebox[14cm][l]{$\displaystyle\rr\ge 0\,\colon
\qquad\qquad\qquad
(1+\O)(\nabla r,\nabla u^+)=
-\frac{(\nabla r)^2}{(\nabla r)^2 +2\beta(v)}\;,$}
\end{equation}
\begin{equation}
\makebox[14cm][l]{$\displaystyle\rr\ge 0\,\colon
\qquad\qquad\qquad
(1+\O)(\nabla v,\nabla u^+)=
-\frac{2}{(\nabla r)^2 +2\beta(v)}\;.$}
\end{equation}
Eqs. (8.8)-(8.13) generalize the respective classical formulae.

The result (8.8) is a specific case of the following result 
for the integral along an outgoing light ray:
\begin{eqnarray}
\makebox[14cm][l]{$\displaystyle\rr\ge 0\,\colon
\quad\;
\int\limits^r_{\rH}\left.\frac{dr}{(\nabla r)^2}f
\right|_{u=\mbox{\scriptsize const.}}=
f_{\mbox{\tiny AH}}(u)
\frac{\rH (u)}{\BH (u)}\ln\frac{1}{2\beta(u)}
+f\frac{r}{B}\ln\left((\nabla r)^2 +2\beta(v)\right)$}\nonumber\\
-\int\limits_0^{(\nabla r)^2}d(\nabla r)^2\left(\ln(\nabla r)^2\right)
\frac{d\hphantom{(\nabla r)^2}}{d(\nabla r)^2}\left(f\left.\frac{r}{B}
\right|_{u=\mbox{\scriptsize const.}}\right)\;.
\end{eqnarray}
In the integral remaining on the right-hand side, one may insert 
the explicit expressions (6.39) and (6.41) for $(\nabla r)^2$ 
and $B$. The
result (8.14) is valid up to $\O$ for any function $f$ that
possesses the properties
\begin{equation}
f=O(1)\;,\qquad
\left.\frac{d\hphantom{(\nabla r)^2}}{d(\nabla r)^2}
\,f\right|_{u=\mbox{\scriptsize const.}}=O(1)
\end{equation}
including at the AH. The property of derivative is possessed,
in particular, by any function that depends on $v$ only
through the arguments
\begin{equation}
f=f(r, E)
\end{equation}
and has bounded derivatives with respect to these arguments.
Indeed, by eqs. (4.37), (4.68), and (1.13),
\begin{equation}
\left.\frac{d\hphantom{(\nabla r)^2}}{d(\nabla r)^2} 
\,r\right|_{u=\mbox{\scriptsize const.}}=
r\,\frac{(\nabla r)^2}{B(\nabla r)^2 -4A}\;,
\end{equation}
\begin{equation}
\left.\frac{d\hphantom{(\nabla r)^2}}{d(\nabla r)^2}
\,E\right|_{u=\mbox{\scriptsize const.}}=
r\,\frac{2A}{B(\nabla r)^2 -4A}-
\frac{1}{2}\,r\,\frac{(\nabla r)^2}{B(\nabla r)^2 -4A}
\left(B-1+(\nabla r)^2\right)\;.
\end{equation}
For $0\le\rr<1-|\O|$, these derivatives are bounded owing
to the negativity of $A$.

For the proof of eq. (8.14), first calculate using
expressions (6.39) and (6.41)
\begin{eqnarray}
\makebox[14cm][l]{$\displaystyle\rr\ge\rr\Bigl|_L\,\colon
\qquad
\int\limits^r_{r\bigl|_L}\left.\frac{dr}{(\nabla r)^2}
f\right|_{u=\mbox{\scriptsize const.}}
=f\frac{r}{B}\ln(\nabla r)^2
-\left.\left(f\frac{r}{B}\ln(\nabla r)^2\right)\right|_L$}
\nonumber\\
-\int\limits^{(\nabla r)^2}_{(\nabla r)^2\bigl|_L}
d(\nabla r)^2\left(\ln(\nabla r)^2\right)
\frac{d\hphantom{(\nabla r)^2}}{d(\nabla r)^2}
\left(f\left.\frac{r}{B}\right|_{u=\mbox{\scriptsize const.}}
\right)\;.
\end{eqnarray}
By conditions (8.15) and the boundedness of the derivative
(8.17), one has
\begin{equation}
\int\limits_0^{(\nabla r)^2\bigl|_L}
d(\nabla r)^2\left(\ln(\nabla r)^2\right)
\frac{d\hphantom{(\nabla r)^2}}{d(\nabla r)^2}
\left(f\left.\frac{r}{B}\right|_{u=\mbox{\scriptsize const.}}
\right)=\O\;.
\end{equation}
Therefore, the lower limit of the integral on the right-hand
side of eq. (8.19) can be shifted to zero. By conditions
(8.15) one has also
\begin{equation}
\makebox[14cm][l]{$\displaystyle\rr=|\O|\,\colon
\qquad\qquad
\int\limits^r_{\rH}\left.\frac{dr}{(\nabla r)^2}
f\right|_{u=\mbox{\scriptsize const.}}
=f_{\mbox{\tiny AH}}(u)
\int\limits^r_{\rH}
\left.\frac{dr}{(\nabla r)^2}
\right|_{u=\mbox{\scriptsize const.}}+\O\;.$}
\end{equation}
The latter integral is the solution of eq. (4.37), and it has 
already been calculated in eq. (8.5):
\begin{equation}
\makebox[14cm][l]{$\displaystyle\rr=|\O|\,\colon
\qquad
\int\limits^r_{\rH}\left.\frac{dr}{(\nabla r)^2}
f\right|_{u=\mbox{\scriptsize const.}}
=f_{\mbox{\tiny AH}}(u)
\frac{\rH (u)}{\BH (u)}\ln
\frac{(\nabla r)^2+2\beta(v)}{2\beta(u)}+\O\;.$}
\end{equation}
Hence
\begin{equation}
\int\limits_{\rH}^{r\bigl|_L}\left.\frac{dr}{(\nabla r)^2}
f\right|_{u=\mbox{\scriptsize const.}}
=f_{\mbox{\tiny AH}}(u)
\frac{\rH (u)}{\BH (u)}\ln
\frac{(\nabla r)^2\Bigl|_L}{2\beta(u)}+\O\;.
\end{equation}
Combining eqs. (8.19) and (8.23) yields the result
\begin{eqnarray}
&\makebox[15cm][l]{$\displaystyle\rr\ge\rr\Bigl|_L\,\colon
\qquad
\int\limits^r_{\rH}\left.\frac{dr}{(\nabla r)^2}
f\right|_{u=\mbox{\scriptsize const.}}
=f_{\mbox{\tiny AH}}(u)
\frac{\rH (u)}{\BH (u)}\ln\frac{1}{2\beta(u)}
+f\frac{r}{B}\ln(\nabla r)^2$}&\nonumber\\
&{\displaystyle\hphantom{\qquad\qquad\qquad\qquad\qquad}
-\int\limits_0^{(\nabla r)^2}d(\nabla r)^2
\left(\ln(\nabla r)^2\right)
\frac{d\hphantom{(\nabla r)^2}}{d(\nabla r)^2}
\left(f\left.\frac{r}{B}\right|_{u=\mbox{\scriptsize const.}}
\right)+\O\;.}&
\end{eqnarray}
Equations (8.22) and (8.24) calculate the integral in 
the (overlapping) regions of strong field and weak field 
respectively. Using these equations, one can check that, 
up to $\O$, eq. (8.14) is valid globally.

The virtue of the global solutions (8.8) and (8.14) is in the fact
that they calculate the respective quantities outside the
region of strong field. In the range
\begin{equation}
|\O|<\rr<1-|\O|\;,
\end{equation}
all terms of expression (8.14) except the first one are $O(1)$ and 
are negligible:
\begin{equation}
\int\limits^r_{\rH}\left.\frac{dr}{(\nabla r)^2}
f\right|_{u=\mbox{\scriptsize const.}}
=f_{\mbox{\tiny AH}}(u)
\frac{\rH (u)}{\BH (u)}\ln\frac{1}{2\beta(u)}
+MO(1)\;.
\end{equation}
Similarly, in this range expression (8.8) is
\begin{equation}
v-\vH (u)=\frac{4\EH (u)}{\BH (u)}\ln\frac{1}{2\beta(u)}
+MO(1)\;.
\end{equation}
The large term proportional to $\ln\beta(u)$ emerges as a
contribution of the strong-field region and dominates in
these expressions.

Eq. (8.27) gives the life-time of the "instantaneous" black
hole, i.e., of what appears as a black hole at each instant
of evaporation. Suppose that some falling observer hits the
AH at a given value of $u$. Then how much later should another
observer fall to discover that, at this value of $u$, there
is no more black hole? The answer is in eq. (8.27). In 
particular, taken at the tangent ray $u=u_0$, eq. (8.27) gives
the life-time of the "classical" black hole, i.e., the one
that forms initially as a result of the collapse but then 
destroys itself by evaporation. In the classical geometry,
this life-time is infinite. The earliest observer to discover
that, at $u=u_0$, there is no more black hole is also the
first to discover that the geometry is no more classical. Hence
\begin{equation}
v_{\mbox{\scriptsize crit}}-v_0=4M\ln\frac{1}{2\beta_0}
+O(M)\;,\qquad
\beta_0=-\left.
\frac{d\rH^{\hphantom{+}}}{du^+_{\hphantom{\tiny AH}}}
\right|_{u=u_0}\;.
\end{equation}
This is the critical value of $v$ provided for by the 
correspondence principle, eq. (2.4).
}

\newpage

{\renewcommand{\theequation}{9.\arabic{equation}}

\begin{center}
\section{\bf    Summary}
\end{center}

$$ $$

The above is all that kinematics can say. It expresses the
metric in the semiclassical region through two Bondi charges
${\cal M}(u)$ and $Q^2(u)$, and this expression is valid
as long as the Bondi charges satisfy the conditions
\begin{equation}
{\cal M}>|\O|\;,
\end{equation}
\begin{equation}
\frac{d{\cal M}}{du^+}=\O\;,\qquad
\frac{dQ^2}{du^+}=\O\;,\qquad
-|\O|\le\frac{Q^2}{{\cal M}^2}<1-|\O|\;.
\end{equation}
Their fulfillment is a verifiable assumption. On the basis
of this assumption, the Bondi charges can be calculated.
Thereby the bound in $u^+$ to the validity of these conditions
will be established. For sufficiently early $u^+$, they
hold deliberately but it may well be the case that there
is a finite value of $u^+$ beyond which they are no longer
valid. If condition (9.1) ceases being valid, then the limit 
of validity of semiclassical theory is reached. However, it
may also be the case that, at some value of $u^+$, conditions
(9.2) cease being valid while condition (9.1) still holds.
It is only then and beyond this value of $u^+$ that the
present approach will fail. The failure will possibly signify
that there are some other semiclassical effects in the problem,
different from and additional to the Hawking effect.

The Planck constant makes no appearance in the present study.
It will appear at the next stage of the calculation. Given
the metric in the semiclassical region, one can use any of the
semiclassical techniques to calculate $\T$ at $\I$ and,
thereby, express the Bondi charges through themselves. The
result will be closed equations for the data functions.
In particular, the constant $\beta_0$ in eq. (8.28) will be
related to the Hawking luminosity of the black hole. Kinematics
reduces the problem in functions of two variables to a problem
in functions of one variable. The latter problem is a subject
of the quantum dynamics. 
}

\newpage
\appendix
\setcounter{equation}{0}

{\renewcommand{\theequation}{A.\arabic{equation}}

\begin{center}
\section*{\bf Appendix. Bound on the first derivative of the\\ 
curvature $A$}
\end{center}

$$ $$

Denote by prime the derivative $M(d/dr)$ along $\v$ Then
eq. (3.10) can be written as
\begin{equation}
\hphantom{d=O(1)\;.\qquad}
|A''|<d\;,\qquad d=O(1)\;.
\end{equation}
Eq. (4.29) implies that, in the region of strong field, $A$
satisfies the condition
\begin{equation}
\hphantom{{\bf A}={\bf A}(v)=\O\;.\qquad}
|A|<{\bf A}\;,\qquad {\bf A}={\bf A}(v)=\O\;.
\end{equation}
By eqs. (4.23), (4.20), and (4.1) this condition holds
on the line $\v$ in the interval of $r$ having the length
\begin{equation}
\hphantom{\forall\O\;.\qquad}
\Delta r=M\O\;,\qquad\forall\O\;.
\end{equation}
For the derivation below, it suffices that the interval of
validity of condition (A.2) have the length
\begin{equation}
\Delta r>2M\sqrt{{\bf A}}\;,
\end{equation}
which is, of course, secured by eq. (A.3). It will then be
proved that
\begin{equation}
\hphantom{c=O(1)\;.\qquad}
|A'|<c\sqrt{{\bf A}}\;,\qquad c=O(1)\;.
\end{equation}

The assertion to be proved is a corollary of a general lemma.
Given bounds on a function and its second derivative, the
lemma establishes a bound on the first derivative provided
that the given bounds hold in a sufficiently large interval
of the argument. For given $d$ in eq. (A.1) and ${\bf A}$
in eq. (A.2), this interval of the argument should exceed
$2\sqrt{{\bf A}}$, and then
\begin{equation}
|A'|<(1+2d)\sqrt{{\bf A}}\;.
\end{equation}

For the proof of the lemma, consider any point in the interval
of validity of conditions (A.1) and (A.2). Call it point 1.
Suppose that
\begin{equation}
|A'_1|>\sqrt{{\bf A}}\;,
\end{equation}
and let, for definiteness, $A'_1$ be positive. Consider any
subinterval containing point 1:
\begin{equation}
\overline{r}\le r_1\le\overline{\overline{r}}
\end{equation}
and having the length
\begin{equation}
\overline{\overline{r}}-\overline{r}=2M\sqrt{{\bf A}}\;.
\end{equation}
It will be proved that, in this subinterval, there is a point 2
at which
\begin{equation}
\hphantom{r\le r_2\le r\;.\qquad}
A'_2=\sqrt{{\bf A}}\;,\qquad 
\overline{r}\le r_2\le\overline{\overline{r}}\;.
\end{equation}
Indeed, if there is no such point, then
\begin{equation}
\hphantom{r\le r\le r\qquad}
A'>\sqrt{{\bf A}}\;,\qquad
\overline{r}\le r\le\overline{\overline{r}}\;,
\end{equation}
and one obtains using eq. (A.2):
\begin{equation}
2{\bf A}>|\overline{\overline{A}}|+|\overline{A}|
\ge|\overline{\overline{A}}-\overline{A}|
=\overline{\overline{A}}-\overline{A}=
\frac{1}{M}\int_{\overline{r}}^{\overline{\overline{r}}}A'dr
>\sqrt{{\bf A}}\,
\frac{\overline{\overline{r}}-\overline{r}}{M}\;.
\end{equation}
Hence
\begin{equation}
\overline{\overline{r}}-\overline{r}<2M\sqrt{{\bf A}}
\end{equation}
which is at variance with eq. (A.9).

Thus the point 2 exists, and
\begin{equation}
|r_1-r_2|\le 2M\sqrt{{\bf A}}\;.
\end{equation}
Then one obtains using eq. (A.1):
\begin{equation}
A'_1-\sqrt{{\bf A}}=|A'_1-A'_2|=\frac{1}{M}
\left|\int^{r_1}_{r_2}A''dr\right|
<d\,\frac{|r_1-r_2|}{M}\le 2d\sqrt{{\bf A}}\;.
\end{equation}
Having conducted the same consideration for negative $A'_1$,
one concludes that, if the assumption (A.7) is true, then
\begin{equation}
|A'_1|<(1+2d)\sqrt{{\bf A}}\;.
\end{equation}
But, if the assumption (A.7) is not true, eq. (A.16) is true
all the more. This proves the result (A.6) for any point in
the region of validity of conditions (A.1) and (A.2).
}

\newpage

\begin{center}
\section*{\bf Acknowledgments}
\end{center}

$$ $$

The present work was supported by the Italian Ministry for
Foreign Affairs, and the Ministry of Education of Japan.
Essential parts of it have been done during the author's
stays at the Yukawa Institute for Theoretical Physics, and
at the University of Naples by the invitation of Centro Volta.
The author is especially grateful for the hospitality and
care to Luigi Cappiello and Roberto Pettorino in Naples,
and Masao Ninomiya and Mihoko Nojiri in Kyoto.

\newpage

\begin{center}
\section*{\bf References}
\end{center}

$$ $$

\begin{enumerate}
\item S.W. Hawking, Commun. math. Phys.
43 (1975) 199.
\item See A.G. Mirzabekian and G.A. Vilkovisky, Ann. Phys. 270
(1998) 391, and references therein.
\item V.P. Frolov and I.D. Novikov, {\it Black Hole Physics}, Kluwer,
Dordrecht, 1998.
\item V.P. Frolov and G.A. Vilkovisky, in {\it Proc. 2nd Marcel
Grossmann Meeting on General Relativity, Trieste, 1979} (R. Ruffini, 
Ed.), p. 455, North-Holland, Amsterdam, 1982.
\item V.P. Frolov and G.A. Vilkovisky, Phys. Lett. B 106 (1981)
307; in {\it Proc. 2nd Seminar on Quantum Gravity, Moscow, 1981}
(M.A. Markov and P.C. West, Eds.), p. 267, Plenum, London, 1983. 
\end{enumerate}

\newpage

\begin{center}
\section*{\bf Figure captions}
\end{center}

$$ $$

\begin{itemize}
\item[Fig.1.] Penrose diagram for the semiclassical region.
The bold curve is the apparent horizon. The light lines
are level lines of the advanced time $v$ and retarded time $u$.
The points 0 to 6 are referred to in the text.
\item[Fig.2.] Completed diagram of Fig. 1. The bold line
$\u$ is the event horizon. The point I on the apparent horizon
is the maximum of $-(d\EH/du)$. The point II is the maximum of
$-\alpha(d\EH/dv)$. The broken line is the line of extrema
of $(\nabla r)^2$. The light solid curve is $\partial_v E=0$.
\end{itemize}

\end{document}